\documentclass[preprint,12pt,3p]{elsarticle}
\usepackage{amssymb}
\usepackage{amsmath}
\usepackage{mathtools}
\usepackage{breqn}
\usepackage{multirow}
\usepackage{tablefootnote}
\usepackage{footnote}
\usepackage{caption}
\usepackage{tabularx}
\usepackage{longtable}
\usepackage{float}
\usepackage{adjustbox}
\usepackage{xcolor}

\usepackage{graphics}
\usepackage[hyphens]{url}
\usepackage[hidelinks]{hyperref}
\hypersetup{breaklinks=true}

\begin{document}

\begin{frontmatter}

%   \title{Blockchain, Fog computing, and AI for Secure, Efficient and Smart IoT-healthcare monitoring system}
%   \title{IoT and AI Enabled  Secure  Remote Patient Monitoring Application}
  \title{A Lightweight Blockchain and Fog-enabled  Secure Remote Patient Monitoring System}
% \tnotetext[label0]{This is only an example}

\author[label2,labelSA,labelMA]{Omar Cheikhrouhou\corref{cor1}}

\address[label2]{CES Laboratory, University of Sfax, Tunisia}
\ead{omar.cheikhrouhou@isetsf.rnu.tn}

\author[label3]{Khaleel Mershad}
\address[label3]{Computer Science and Mathematics Department, Lebanese American University, Beirut, Lebanon}
\ead{khaleel.mershad@lau.edu.lb}

\author[label4]{Faisal Jamil}
\address[label4]{Department of Computer Engineering, Jeju National University, Korea}
\ead{faisal@jejunu.ac.kr}

\author[labelred]{Redowan Mahmud}
\address[labelred]{School of Electrical Engineering, Computing and Mathematical Sciences, Curtin University, Perth, Australia}
\ead{mdredowan.mahmud@curtin.edu.au}

\author[labelSA]{Anis Koubaa}
\address[labelSA]{Robotics and Internet of Things Lab, Prince Sultan University, Riyadh, Saudi Arabia.}
\ead{akoubaa@psu.edu.sa}

\author[label5]{Sanaz Rahimi Moosavi}
\address[label5]{California State University, Dominguez Hills (CSUDH)}
\ead{srahimimoosavi@csudh.edu}

\address[labelMA]{ISIMA, Mahdia, University of Monastir, Tunisia\fnref{label4}}

\cortext[cor1]{I am corresponding author}
% \fntext[label3]{I also want to inform about\ldots}
% \fntext[label4]{Small city}

% \ead[url]{author-one-homepage.com}

\begin{abstract}
IoT has enabled the rapid growth of smart remote healthcare applications. These IoT-based remote healthcare applications deliver fast and preventive medical services to patients at risk or with chronic diseases. However, ensuring data security and patient privacy while exchanging sensitive medical data among medical IoT devices is still a significant concern in remote healthcare applications. Altered or corrupted medical data may cause wrong treatment and create grave health issues for patients. Moreover, current remote medical applications' efficiency and response time need to be addressed and improved. Considering the need for secure and efficient patient care,  this paper proposes a lightweight Blockchain-based and Fog-enabled remote patient monitoring system that provides a high level of security and efficient response time. Simulation results and security analysis show that the proposed lightweight blockchain architecture fits the resource-constrained IoT devices well and is secure against attacks. Moreover, the augmentation of Fog computing improved the responsiveness of the remote patient monitoring system by 40\%.

\end{abstract}

\begin{keyword}
 IoT, Healthcare monitoring, Lightweight Blockchain, Fog computing, consensus protocol.
\end{keyword}

\end{frontmatter}

% \linenumbers

\section{Introduction}
Healthcare IoT networks are evolving from centralized to distributed systems to connect with each other to provide patients with high-quality healthcare. According to predictions, the current hospital-centered healthcare monitoring systems will develop first to hospital–home-balanced in 2025 and then ultimately to home-centered in 2030 \cite{rahmani2018exploiting}. New system architectures, technologies, and computing paradigms are needed to realize such evolution, specifically in the Healthcare Internet of Things (HIoT) \cite{zaabar2021secure}. 
Emerging technologies like IoT, blockchain, and artificial intelligence have made deploying smart remote patient monitoring systems a fact. 
Indeed, IoT devices permit them to sense and monitor patients' physiological parameters, hence exempting them from a long waiting queue at a doctor's visit. All necessary physiological parameters needed by doctors can be sensed by the biomedical IoT devices (also known as the Internet of Medical Things devices) and sent remotely to the doctor, allowing the latter to decide the appropriate treatment for the patient \cite{cheikhrouhou2021one}.

% \color{red}
The evolution of sophisticated security attacks and the rising need for individualized healthcare has made it essential for medical institutions to embrace blockchain technology. The arrival of the blockchain provides solutions to several problems that the healthcare system has been facing for a long time. The growing numbers of healthcare data breaches, patient privacy violations, counterfeit drugs, and many other issues are major reasons for steering the blockchain market's growth in the healthcare industry. In general, the blockchain brings a large number of opportunities to smart healthcare, which can be summarized as follows:
\begin{itemize}
\item 
Secure access to personal health records: the decentralized blockchain system offers the power of controlling data access to the owner of the data itself. Smart contracts register and authorize users to access the patient's data according to the patient consent policy.
\item 
Patient Consent Management: the fundamental features of the blockchain, such as transparency and immutability, enables healthcare applications to build trust among patients and verify compliance with consent management policies.
\item 
Traceability of remote treatment: the blockchain permits healthcare applications to create immutable and coherent electronic records (EHRs) that can be viewed by all stakeholders. The transparency and consistency of blockchain EHRs aid in tracing the medical history of patients to offer the appropriate treatment.
\item 
Traceability of in-home medical kits and devices: the blockchain provides immutable and transparent record transactions to the ownership and performance of medical kits. Reputation scores of medical devices and kits are saved in the blockchain using smart contracts.
\item 
Reputation-aware specialist referral services: during the treatment of a remote patient, medical referrals and expert suggestions are acquired through smart contracts. Blockchain enables healthcare providers to store these referral documents on an InterPlanetary File System (IPFS) server, such that an IPFS hash of the document is stored securely in the blockchain. The hash prevents the alteration of the stored document and maintains its integrity.
\item 
Automated payments: blockchain provides digitally signed automatic payments to guarantee non-repudiated secure transactions.
\end{itemize}

A complete discussion on the blockchain benefits to smart healthcare applications can be found in \cite{ahmad2021role}.

\color{black}

Ensuring the security of the remote patient monitoring (RPM) system is a must. Since a vulnerability in such a system could enable attackers to steal/modify sensitive information and endanger the patient's life. %Recently, 
The blockchain has emerged as a promising technology that can store and secure assets through a transparent and distributed ledger. In healthcare, where patient data is a critical asset that needs to be securely managed, the blockchain could become the right technology to address this challenge and provide a secure, transparent, and tamper-proof management of patient healthcare data. However, the blockchain is a heavy system requiring much processing and communication. Lightweight IoT devices would face problems if they were to act as full blockchain nodes. Hence, a solution should be adopted to enable IoT devices to participate in the blockchain network without affecting their limited resources. The lightweight blockchain \cite{dai2019Blockchain, mershad2022proact} has been proposed to achieve this purpose. Here, the blockchain architecture and processes are modified to assign light roles to the IoT devices while allowing them to benefit from the blockchain services.

In traditional RPM systems, patient healthcare data is stored in an Electronic Healthcare Record (EHR) and saved in the cloud. Cloud computing provides ubiquitous access to patients' data through a user-centric access control model, where the user chooses which data and to whom he/she should give access. However, a cloud computing system presents the disadvantages of high latency and, therefore, cannot fit critical healthcare application requirements where immediate intervention is needed. More precisely, real-time detection and notification of abnormal situations must be implemented in the context of a heart disease use case. Otherwise, the patient's life will be at risk. 

To overcome the high latency limits of cloud computing and to fit the real-time requirements of most healthcare applications, we propose leveraging fog computing technology in this paper. In our proposed architecture, fog computing will not replace cloud computing but will cooperate via the lightweight blockchain to provide real-time and efficient service. More precisely, we introduce the fog computing layer that will host a lightweight blockchain application with low latency requirements. On the other hand, complex AI algorithms can be executed at the cloud computing layer.

% \color{red}
Currently, smart cities are moving towards adopting blockchain technology in many smart city applications. In healthcare, and especially in remote patient monitoring, the blockchain can change the methods in which the application is executed and managed. Integrating the blockchain allows healthcare managers to guarantee the transparency of public healthcare data and removes the need to apply trust-based mechanisms and systems to achieve this target. In addition, the blockchain guarantees the privacy of patients' personal data through smart contracts. Moreover, the blockchain allows for fast and direct connectivity between healthcare officials, providers, staff, and patients. Issuing blockchain transactions allows these entities to communicate securely via the blockchain without intermediaries. Finally, the blockchain allows healthcare and smart city officials to know the origin and destination of each medical resource. They can also find out how healthcare services are being used without compromising people's privacy.

%deploying an integrated blockchain-fog RPM system enables smart city administrators to reduce the energy consumption of the application by decreasing the heavy load on the healthcare providers' servers and distributing some of that load among the fog nodes which are near the IoT ecosystem.  

% Safe and high-quality healthcare service is of paramount importance to patients. Accordingly, healthcare data security and patients' privacy are important issues that will significantly impact the future success of HealthIIoT [17]. One of the significant issues in the IIoT-based healthcare system is privacy protection.
\color{black}
To sum up, we propose a smart and secure remote patient monitoring system based on three technology pillars: IoT, fog computing, and blockchain. More precisely, the key contributions of this paper are as follows:

% \color{red}
\begin{itemize}
\item We propose the architecture of a smart and secure remote patient monitoring system. The proposed architecture uses IoT for patient vital signs collection and blockchain to guarantee the privacy and security of the patient-collected data. 
\item The efficiency of the proposed architecture is achieved through the introduction of the fog computing layer to provide real-time response and aggregate the patients' collected data.
\item To reduce the heavy demands of traditional blockchain, we modify the blockchain structure to include a local blockchain within the IoT ecosystems and a global chain at the cloud layer. Each IoT ecosystem saves the block headers of all blockchain blocks, the bodies of the blocks of interest to the local chain, and the smart contract functions needed within the local chain. On the other hand, the global chain comprises whole blocks and smart contracts.
\item We propose a lightweight consensus model that enables the fog nodes to participate in the consensus protocol without consuming a lot of processing and energy resources and allows IoT nodes to store only the information they need to verify the legitimacy and integrity of the blockchain data that they obtain from fog nodes and cloud servers. 
\end{itemize}

\color{black}
The remainder of this paper is as follows.  \autoref{sec:related} outlines the existing literature on the remote patient monitoring system using blockchain and Fog Computing. \autoref{sec:proposedarchitect} gives an overview of the proposed remote patient monitoring architecture with its different components.  
\autoref{sec:Blockchain} describes the details of the proposed lightweight blockchain model. 
\autoref{sec:fog} describes the fog computing layer functions and properties. 
The performance evaluation of the system is discussed in \autoref{sec:perfEval}. 
\autoref{sec:security} analyses the security of the proposed system.
Finally, we conclude and give future directions in \autoref{sec:concl}.

\section{Related Work}
\label{sec:related}
As our work is based on three technologies, namely the IoT, fog computing, and blockchain, in this section, we present relevant work that uses one or more of these technologies to deploy a healthcare solution. The discussed works are summarized in \autoref{tab:sumRelWork}.

\begin{center}
\begin{longtable}{|m{0.5 cm}|m {5 cm}|m {3 cm}|m{0.5 cm}|m{0.5 cm}|m{0.5 cm}|m{3 cm}|}
\caption{Summary of related work}
\label{tab:sumRelWork}
% \resizebox{\textwidth}{!}{%
% \begin{tabular}{|m{0.5 cm}|m {3 cm}|m {2 cm}|m{0.5 cm}|m{0.5 cm}|m{0.5 cm}|m{0.5 cm}|m{3 cm}|}
\\ \hline
\multirow{2}{*}{Ref } & \multirow{2}{*} {Contribution} &  \multirow{2}{*}{Use case} &  \multicolumn{3}{c|}{Used Technologies}  & \multirow{2}{*}{Pros(+)/Cons(-) }  \\
\cline{4-6}
 &&& IoT & BC & FC\\
 \hline
 \endhead
 
 \hline \multicolumn{7}{|r|}{{Continued on next page}} \\ \hline
\endfoot

\multicolumn{7}{|l|}{{BC: Blockchain, FC: Fog Computing or any cloud computing related technologies}} \\ \hline
\endlastfoot
 
 \cite{hossain2016cloud} &   Cloud based remote health monitoring system with signal watermarking    &  ECG-based health monitoring        &  \checkmark    &      & \checkmark   &    +Providing signal authentication using watermarking     \\   \hline
%  Hich: Hierarchical fog-assisted computing architecture for healthcare IoT
 \cite{azimi2017hich}& A hierarchical fog-computing-assisted architecture for IoT health monitoring system &  Arrhythmia detection     &  \checkmark    &      &  \checkmark   &     + Map the IBM’s MAPE-K computing model to the healthcare application    \\   \hline
 
 \cite{rahmani2018exploiting} &  They developed a smart e-Health gateway localized at the edge.        &     Heart disease     & \checkmark    &            &\checkmark  &  +Full-system implementation   \\ \hline

\cite{pagan2018power}&  Improved the energy consumption of sensor nodes during data transmission and processing.  
&     Migraine disease     & \checkmark    &            &\checkmark  & +Energy consumption reduction   \\ \hline

\cite{pace2018edge}& An Edge-Based Architecture for IoT-Healthcare application. &  Detect high-stress conditions for workers and athletes.     & \checkmark    &            &\checkmark  & -security issues are not addressed.   \\ \hline

\cite{juneja2018leveraging}& Used retraining of SDA in the testing phase of arrhythmia classification to add or merge features in the anomaly detector + Blockchain for access control &    arrhythmia classification    &     &    \checkmark         &  & +High accuracy   \\ \hline

\cite{griggs2018healthcare}& Used blockchain to secure remote patient monitoring &    General       &    \checkmark        &  \checkmark  &  & 
% -Time issue -Key management issue
% \begin{itemize}
\item  -Time issue  
\item  -Key management issue
% \end{itemize}
 \\ \hline

\cite{ali2020decentralized}    &      Remote health monitoring system using Tor to minimize the latency of the blockchain network.        &  Cardiac Patients. Sleep Apnoea Patients. Epileptic Patients        &  \checkmark    &  \checkmark    &    &  -The accuracy of the system is not tested.      \\    \hline

% \cite{ali2020smart}    &   A  heart disease prediction system based on ensemble deep learning and feature fusion          &     Heart disease prediction    & \checkmark    &             & &    -The network architecture and the database location are not precise.   \\ \hline

\cite{tuli2020healthfog}    &   HealthFog: A  heart disease analysis system based on ensemble deep learning and using integrated IoT and Fog computing          &     Heart disease     & \checkmark    &             &\checkmark & -Security issue are not addressed.   \\ \hline

\cite{farahani2020towards}&  They developed an intelligent e-Health architecture integrating AI, IoT, and cloud computing.        &     ECG-based arrhythmia detection     & \checkmark    &               &\checkmark  &  +Hardware implementation of AI algorithms   \\ \hline

\cite{garcia2020accelerating}&  An IoT and fog computing architecture with parallelization and core allocation capabilities to accelerate healthcare processor-intensive services         &     ECG-based arrhythmia detection     & \checkmark    &                &\checkmark  &  +Response Time was improved.   \\ \hline

\cite{wang2022dag} & Lightweight identity management and access control scheme for IoT devices using IOTA. & General &\checkmark & \checkmark & & -Does not support smart contracts\\ \hline

\cite{hossein2021bchealth} & Blockchain based architecture to provide patient centric data access & Healthcare & \checkmark & \checkmark  & & + Use clustering techniques to improve system scalability\\ \hline

% \cite{ref12} & Implements Echo State Network with Reservoir computing algorithm & ECG Arrhythmia classification & & & \checkmark & & + High accuracy \\ \hline

% \cite{ref13} & Implements Parallel Convolution Neural Network & ECG Arrhythmia classification & & & \checkmark & & + The network is optimized using Genetic Algorithm \\ \hline

% \cite{ref14} & Implements Deep residual network with skip connections & ECG Arrhythmia classification & & & \checkmark & & + High accuracy \\ \hline

% \cite{ref15} & Proposed Recurrent Network with LSTM cells & ECG Arrhythmia classification & & & \checkmark & & + Focal loss is applied to resolve imbalance beat data\\ \hline

% \cite{ref16} & Implements Deep RNN with one-hot encoding layer & ECG Arrhythmia classification & & & \checkmark & & + High accuracy\\ \hline

%     \multicolumn{7}{l|}{\tinysize{ BC: Blockchain, }}
% \end{tabular}}
% \left
% \caption*{BC: Blockchain, FC: Fog Computing or any cloud computing related technologies}
\end{longtable}
\end{center}

Hossain et al. \cite{hossain2016cloud} proposed a cloud-based architecture for ECG signal monitoring. To authenticate the captured ECG signal, the authors add a watermark that will be checked on the cloud side. Moreover, the authors proposed additional services, including ECG signal enhancement, classification, and analysis.
Azimi et al., \cite{azimi2017hich} proposed a hierarchical computing architecture leveraging fog and cloud computing technologies. The authors proposed a methodology to partition the existing machine learning methods for fog-enabled healthcare IoT systems.  
The authors in \cite{rahmani2018exploiting} developed a smart e-Health gateway localized at the edge to provide several functions, including local storage, real-time local data processing, embedded data mining, etc. 
By releasing the small IoT devices from these functions, a considerable amount of energy can be saved by outsourcing some loads from sensor nodes to these smart gateways.

In \cite{pagan2018power}, the authors also proposed the integration of the IoT and cloud computing technologies to predict migraine disease. The authors' main contributions are the design of low-power techniques in the radio and data processing for the sensor nodes. Moreover, the authors proposed workload-balancing policies for cloud computing servers.

The authors in \cite{pace2018edge} proposed \textit{BodyEdge}: an edge-based architecture for IoT-healthcare applications. The system was implemented in two examples of edge gateway: Raspberry Pi3 and Zotac CI540 NANO Pc, and its performance was compared to cloud systems. Moreover, as a validation example, the authors have implemented the system to detect high-stress conditions for users in two different scenarios, namely Workers in a factory and Athletes training in a fitness center. 
In \cite{juneja2018leveraging}, authors used blockchain to secure access control to patient EHR. The authors proposed to store patient data in an off-chain database to overcome the storage constraint of the blockchain. 
% Regarding the deep learning architecture, the authors proposed to use Stacked Denoising Autoencoders (SDA). 
% Blockchain is a decentralized distributed ledger that
% secures transactions with cryptography. It is proposed as an
% access control manager to securely store and access data
% required by the classifier during retraining in real-time from
% an external data storage.

The authors in \cite{griggs2018healthcare} used a consortium blockchain based on the IBM Hyperledger platform to secure remote patient monitoring. In their proposed system, sensors interact with a gateway (such as a mobile phone) that implements smart contracts for data analysis and sends essential notifications to patients and healthcare providers. The blockchain was used to securely log transactions (such as data reads and doctor's commands). However, patient data was stored on a local database. They have proved that blockchain could be used to resolve security concerns about the transfer and logging of data transactions in an IoT healthcare system. The limitation rests in perfecting the time of the transmission of the aggregated data sent by the gateway to the blockchain nodes.

The authors in \cite{ali2020decentralized} proposed a decentralized peer-to-peer remote health monitoring system. 
The proposed architecture uses Tor hidden services for off-chain data delivery between patients and doctors. 
The authors in \cite{tuli2020healthfog} proposed HealthFog, a Fog-based healthcare system that integrates Edge computing and IoT. Their work was motivated by latency-sensitive healthcare applications, especially deep learning-based algorithms. The proposed system was validated for a health disease use case.
% //include healthfog related work .....

In \cite{farahani2020towards}, an AI-driven e-health solution was proposed. The solution integrates IoT with cloud computing.
The key difference of this solution is the distribution of the AI intelligence across the three architecture layers, namely: the Device layer, Fog layer, and Cloud layer.
Moreover, the authors proposed hardware implementation of the SVM, ANN, and CNN algorithms using digital circuits.
The authors in \cite{garcia2020accelerating} proposed a framework for accelerating the response to remote patients requiring the execution of smart eHealth services. Their proposed framework supports distributed offloading to fog servers and multicore processors' capacity to accelerate its execution.

The authors in \cite{wang2022dag} proposed a blockchain-based lightweight authentication and authorization scheme
for IoT devices. The proposed scheme uses distributed ledger technology IOTA to design a lightweight and scalable mechanism for identity management and access control of IoT devices. However, this solution did not support smart contacts.
The authors in \cite{hossein2021bchealth} proposed a blockchain-based architecture that enables data owners to define their desired access policies over their privacy-sensitive healthcare data. The architecture used two separate chains; one for storing data transactions and one for storing access policies.

Several lightweight blockchain architectures have been proposed in the literature \cite{xie2021eclb, honar2021multi, yao2022accident, mershad2020blockchain, shahid2019sensor, sunny2020towards}. For example, the ECLB protocol in \cite{xie2021eclb} saves the full blockchain on edge nodes, while the IoT nodes store what the authors call the fragmented ledger structure, which contains the block headers and some of the transactions in each block that are needed by the lightweight node. A multi-layer blockchain model is proposed in \cite{honar2021multi}. The blockchain network is divided into three layers. At the first layer, ordinary IoT nodes are divided into clusters. At the second layer, IoT cluster heads (CHs) store the local (i.e., cluster) copy of the BC. The cellular base stations (BSs) store the full global BC at the third layer. Nodes in Layers 2 and 3 collaborate to create new blocks and execute the consensus algorithm. On the other hand, some IoT nodes at layer one can be peers that maintain a copy of the local BC and act as transaction endorsers or committers, while CHs and BSs act as Hyperledger orderers who order transactions and create blocks.

% \color{red}{
From the study of the existing work, we note that several works address only the performance and energy aspect of their proposed RPMs by adding the fog layer that manages the computing and data processing tasks \cite{hossain2016cloud,azimi2017hich,rahmani2018exploiting,pagan2018power,pace2018edge,tuli2020healthfog,farahani2020towards,garcia2020accelerating}.
However, these schemes did not address the security aspects, and therefore, they are vulnerable to attacks.
Other schemes such as \cite{juneja2018leveraging,griggs2018healthcare,ali2020decentralized,wang2022dag,hossein2021bchealth}, addressed the security aspect by adopting the blockchain technology, however, they used classical blockchain platforms and architectures that cannot fit the resource-constrained IoT devices. 
In this paper, we leverage the fog computing layer not only to improve the performance of the RPM system but also to lighten the load of blockchain technology. More precisely, the fog layer permits the proposal of a lightweight blockchain architecture that provides security services adaptable to resources constrained IoT devices. Moreover, the proposed consensus mechanism frees the architecture from the burden of classical consensus algorithms such as PoW or PoS \cite{zaabar2021healthblock,mershad2021proof}.

\color{black}
%
%\autoref{tab:sumRelWork} summarizes the previously discussed related work.
\section{The Proposed RPM System Overview}
\label{sec:proposedarchitect}
% //here we describe the proposed architecture based on IoT Blockchain 
This section gives an overview of the proposed RPM system. It highlights its three-layer architecture and the different communication interaction between the components.
\subsection{RPM System Architecture}
The proposed remote patient monitoring system is a three-layer architecture as shown in  \autoref{fig:rpmArchitecture}, and which are:

\begin{figure}
\centering
\includegraphics[width=0.9\textwidth]{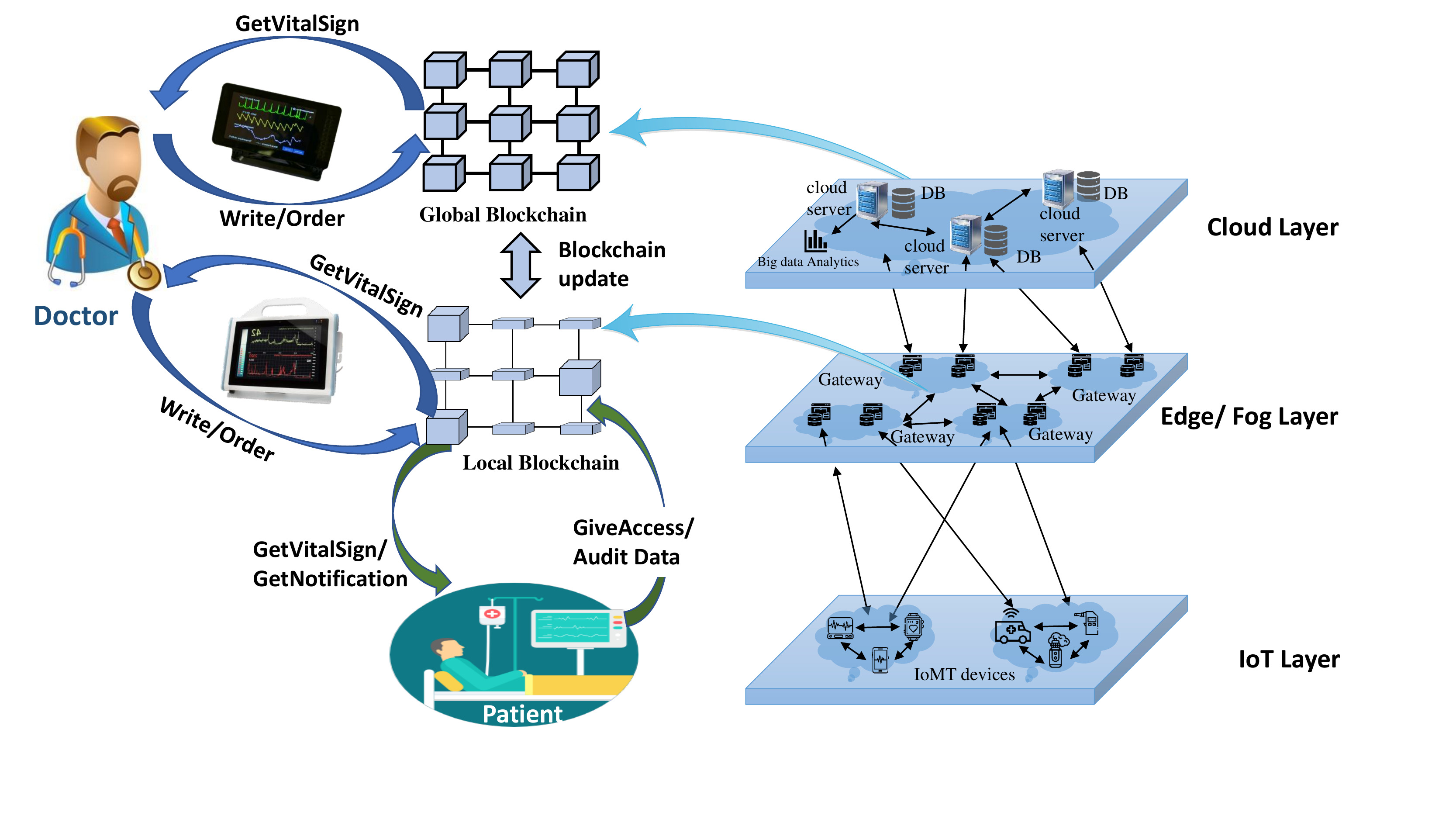}
\caption{The three-layer architecture of the proposed remote patient monitoring system}
\label{fig:rpmArchitecture}
\end{figure}

\begin{itemize}
\item \textbf{The IoT devices layer}: This layer, composed of biomedical sensor nodes, wearable sensor nodes,  and IoT medical devices, is responsible for collecting the vital signs of the monitored patient. These sensor readouts are collected continuously; however, their transmission to the gateway node located at the fog layer can be done periodically. The transmission period depends on the nature of the vital sign and generally is determined by the patient supervising doctors. 
\item \textbf{The Fog Computing layer}: This layer is responsible for the lightweight processing of vital signs received from the IoT layer. For example, suppose the monitored vital sign exceeds a specific threshold. In that case, an alert message will be triggered and sent to the patient and the supervising doctor to make the right decision. Moreover, the fog layer first decides which data needs to be recorded in the blockchain network and then interacts with this latter. The fog layer will also aggregate continuous sensed data before sending it to the cloud server for permanent storage and data analytics.
Additionally, the fog layer contains IoT gateways that include the local blockchain network, which is a subset of the global blockchain network (please refer to \autoref{sec:fog} for full description). In the proposed system, the fog computing module consists of many geographical intelligent gateways, that is, forming the fog. Each gateway supports different protocols for communication and serves as a point of contact between the sensor network and the cloud. It collects data from different sub-networks, translates protocols, and offers other higher-level services, including filters, data aggregation, analysis, and so on. The fog computing layer extends cloud computing to the edge of the network and its facilities. From cloud to end users/devices, the fog recognizes real-time interaction, dense geographical distribution, heterogeneity, accessibility support, pre-processing interoperability along with cloud interaction \cite{Sanaz}. This enables latency to be decreased, particularly for real-time applications such as in-house IoT monitoring of patients. The fog reduces contact with the cloud, particularly in the event of a loss of cloud connectivity, where the data is stored locally on these gateways, and patients' data is sent to the cloud when the connection is restored.
\item \textbf{The Cloud Computing layer}: This layer is responsible for permanent data storage and data analytics. Complex AI and deep learning algorithms can be implemented at this layer for data classification, disease detection and prediction, and treatment plan decision. Moreover,  in the proposed architecture, cloud servers play the role of full blockchain nodes that store the full copy of the blockchain and participate in transaction validation, block generation, and consensus. The blockchain records patient data and actions of patients and caregivers and permits the patients to decide to whom they give access to their data. Moreover, blockchain technology contains pieces of code called smart contracts that can be automatically triggered when an event is achieved. These smart contracts are a powerful tool for a remote patient monitoring system as they can trigger an alarm and notify the doctor in an abnormal situation (for example, when the vital sign value exceeds a specific threshold). In addition, the blockchain is used to ensure patient data privacy and the system's security. First, thanks to blockchain technology, the patient will be given an anonymous identity. This permits hiding the real patient's identity; therefore, doctors can treat his/her data privately. Moreover, in our system, we propose to use a private blockchain. This type of blockchain has the advantage of restricting access to users' data to only authorized persons (such as patients, doctors, and caregivers). Furthermore, blockchain architecture permits a patient-centric data management architecture. More precisely, the patient will decide to whom he/she shares data access  (please refer to \autoref{sec:Blockchain} for more details). 

\end{itemize}
\subsection{Communication Models}

The fog layer enables us to control access to IoT devices for medical applications. Each fog node manages and operates a group of medical IoT devices. This layer also interacts with a network of fog nodes allowed by blockchain, which function together on the Internet. All the related smart medical devices are connected with the closest blockchain-enabled fog node, e.g., in an in-house monitoring scenario. 
These blockchain-enabled fog nodes are communicated by IoT nodes and system users for authentication, authorization, and safe communication synchronization. An intelligent contract with a collection of rules can also be established on top of the fog nodes allowed by blockchain. Furthermore, the consensus algorithm is performed to validate the transactions and blocks for those transactions after they are created. Transaction blocks can be exchanged between cloud servers and the blockchain-enabled fog nodes or between the fog nodes to support robust authentication, permission, and distributed secure communication. The proposed solution  mainly includes four forms of communication:

\begin{enumerate} [(1)]
\item \textbf{Medical caregiver-to-fog communication:} Where the end user (e.g., a healthcare provider) is willing to use a particular IoT system, he first sends a request for authentication with a query authentication function specifying the sensor details to the blockchain-enabled fog node. The fog node with the blockchain feature will search for that medical attendant in the available list of approved sensor equipment. A reject message will be given when the user is not allowed to access the requested data. Otherwise, if the user is approved, the blockchain-enabled fog node issues an access token containing Unique Identification (UID) information, length, time of access, blockchain address for the data, user blockchain address, and blockchain address of the fog node that stores the requested data. Notice that every fog node, sensor, and the user has a unique blockchain address.

\item \textbf{Medical sensor-to-fog communication:} The sensor-to-fog correspondence has two principle objectives in our framework. IoT medical services system mainly aims to validate and authorize the clinical sensors. The following goal is to insert a blockchain-enabled fog connected to sensor devices. It helps new sensors to enlist with the mist and ensures that all sensors are recognizable by the blockchain network. In our context, each IoT medical care system has at least one blockchain-enabled fog node close to the entire blockchain network and is used for the enlistment, confirmation, and authorization of IoT medical care gadgets with the same framework. Initially, the gadgets will enroll with their associated blockchain fog node. As an exchange and blocks are made for them, data concerning these gadgets are placed in the blockchain. These blocks would then be transported between the wide range of different blockchain fog nodes. Should a system with a collecting place need confirmation and consent, the associated blockchain-enabled fog node should be given its certifications. The blockchain approves the provided accreditation, and if there are significant requirements, the IoT gadgets for medical care are effectively checked and authorized. If the certification is not valid, the gadget is refused and will not obtain permission to access the blockchain data. 
\item \textbf{Fog-to-fog communication:} The main goal is to synchronize the information associated with IoT medical service confirmation and approval across all blockchain-enabled fog nodes \cite{Khalid}. Several biological or physiological parameters are obtained by medical sensors transmitted by patients. Medical IoT programs should be reliable and diligent in supporting patients moving to a hospital or home. Typically, the mobility support of the medical sensors from the upper layer (i.e., fog layer) should be given so that zero reconfiguration in the sensor layer is essential. The strategic location and distribution of smart gates in the fog layer can be used to provide smooth mobility for medical sensors and relieve processing loads.
Fog-to-fog contact helps patients wander around the hospital wards, ensuring their health monitoring is not disrupted. The patient-free movement provides a high level of medical services using a portable patient monitoring system. Support of mobility for healthcare IoT systems is one of the most critical problems \cite{cite5}. The improvements to patients' quality of life in such programs are essential \cite{Sanaz}. It is important to encourage patients to walk into the hospital/medical facilities knowing that monitoring their well-being is not disrupted. It is necessary to establish self-configuration or transfer mechanisms to ensure safe and successful data transfer between different Medical Sensor Networks (MSNs) \cite{cite6} to achieve ongoing monitoring of patients considering mobility support. For example, when a patient is moving across the clinics, a data transfer mechanism is described as the process of modifying or updating the registration of mobile sensors on its MSN base. Data handover solutions should allow ubiquity when they need to function independently without human interference.

\item \textbf{Medical sensor-to-medical sensor communication:} When two clinical devices are effectively tested and approved (both have a position with a similar system or another one), they may create a safe link to each other and convey information. In a case, for example, where a patient is released from a clinic but still needs to be constantly monitored. The doctors bind the patient's body before he/she leaves the medical center to health tracking devices, including blood pressure monitors, pulse sensors, blood glucose monitoring sensors, etc. These devices sense the patient's blood pressure, heart rate, and glucose level and transmit them through a safe channel to the health workers. These devices can also interact with the patient's intelligent home devices. For example, if the patient's condition becomes severe or a fall is detected, an immediate alarm may automatically be activated. The hospital-related devices must interact to check the availability of hospital beds in a smart city to ensure a correct count. 
The proposed framework provides medical devices with access control in the IoT healthcare system. Under this mechanism, devices can only communicate with recorded and successfully authenticated and certified devices with blockchain-enabled fog nodes. A device not registered in the blockchain cannot authenticate itself or communicate with other devices within the same healthcare ecosystem or external ones. The contact between malicious devices and legitimate devices would also be alleviated.
\end{enumerate}
In what follows, we detail the proposed lightweight blockchain model and the Fog layer functions and properties.
% these different components of our proposed RPM architecture.

% Khaleel: There are three sections here about the fog layer, Data Analytics, and blockchain. However, the model contains three layers. There are no sections for the IoT layer and the Cloud layer. One of the reviewers mentioned this point and considered it a very negative point. We need to add two brief sections of the two missing layers.

% In the proposed architecture, cloud servers play the role of full blockchain nodes that store the full copy of the Blockchain and participate in transaction validation, block generation,
% and consensus. 

\section{The Blockchain Module Description }
\label{sec:Blockchain}

Blockchain technology provides a decentralized, transparent, authenticated platform that applies a consensus-driven approach to facilitate the interactions of multiple entities through the use of a shared ledger. Beyond the financial sector, where much of the initial development is taking place, blockchain has the potential to revolutionize the healthcare system. By providing doctors, patients, researchers, and other healthcare professionals with a mechanism for the controlled exchange of sensitive, permissioned data, blockchain technology can improve data sharing and transparency between clinical and research data systems. Any healthcare organization participating in a blockchain consortium would be able to share medical information, regardless of their native electronic health record system. Blockchain provides significant opportunities for healthcare organizations to deliver more efficacious treatments and diagnoses through increased provider data sharing and potentially safer and more effective remote patient monitoring through advanced technologies such as AI.

\subsection{Blockchain Architecture of Proposed RPM System}

We propose a lightweight blockchain architecture to manage the data storage and retrieval operations in the remote patient monitoring system. 
In our system, we implement a lightweight blockchain architecture that aims at reducing the delay in accessing the cloud by the end users while maintaining the security and immutability of data at all nodes. The blockchain will store all healthcare-related data, such as the IoT sensor readings, lab test results, physicians’ decisions, commands, etc. In addition, the blockchain will comprise transactions that contain management and security-related data, such as nodes’ and users’ registrations, access requests, smart contract results, etc.

% \subsection{Blockchain Architecture}

In the proposed architecture, cloud servers play the role of full blockchain nodes that store the full copy of the blockchain and participate in transaction validation, block generation, and consensus. On the other hand, IoT gateways play the role of light blockchain nodes that store part of the blockchain. In our system, each gateway will be connected to a certain number of IoT networks. For example, an IoT gateway at a patient’s home will connect to a single IoT network that contains the IoT devices that are monitoring the patient. On the other hand, an IoT gateway at a hospital could connect several IoT networks, such as IoT devices, in several patients’ rooms. Here, the IoT devices in a certain room or Lab form a separate IoT subnetwork since the data produced by these devices will be linked together (for example, data related to a specific patient, doctor, lab, etc.).

The IoT gateways and sensors that exist in the same IoT ecosystem (for example, home, hospital, health institution, etc.) form a cluster that store and manage a local blockchain. Each local blockchain is created as part of the full blockchain that is related to the corresponding ecosystem. For example, in a certain patient’s home, a set of IoT devices are connected to an IoT gateway. The devices and gateway form a cluster that store and manage a local blockchain that contains the blockchain blocks related to that home only. In a hospital, several gateways and sets of IoT devices will form a cluster that store and manage the blockchain of the hospital.

\begin{figure}
\centering
\includegraphics[width=1.05\textwidth]{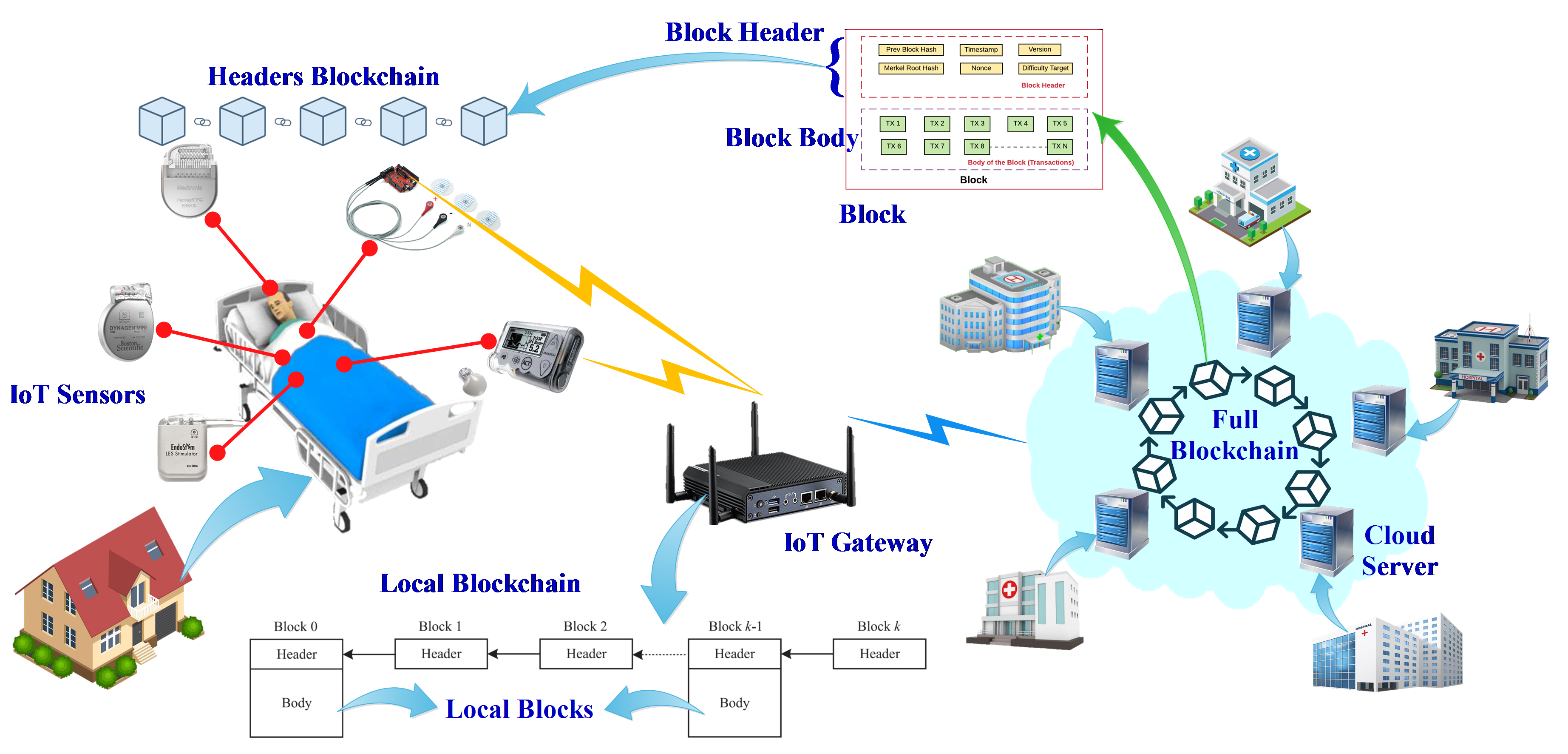}
\caption{The architecture of the proposed blockchain model: the cloud servers store the full blockchain, the IoT gateways save the local blockchain, while the IoT nodes store the block headers.}
\label{fig:BCArchitecture}
\end{figure}

In each cluster, the sensor nodes store only the blocks headers of the full blockchain, while the gateways store the block headers of the full blockchain in addition to the full blocks of the local blockchain (as illustrated in Figure \ref{fig:BCArchitecture}). In addition, to avoid overwhelming the gateways with excessive storage as the blockchain grows, each transaction will have an expiry time after which it becomes obsolete (for example, when the information in the transaction becomes old and is no more relevant). Each gateway saves a data structure that contains, for each transaction, the ID of the block in which the transaction is stored (\textit{BlockID}) and the transaction expiry date (\textit{T\textsubscript{ex}}). The gateway continuously updates the data structure when a transaction expires. In addition, the gateway searches the data structure to detect any block in which all transactions have expired and deletes it. Using this approach allows the gateway to remove old blocks and create room in its storage for new blocks in the local blockchain.   

In the proposed system, IoT nodes continuously generate data and send them to the IoT gateway. In addition, healthcare providers (doctors, nurses, scientists, etc.) send their data (such as prescriptions, sensors' configurations, lab test results, commands to activate actuators, data analytic results, etc.) to the nearest IoT gateway in their institution's IoT cluster. The IoT gateway stores the data it receives in a temporary cache. Each small period (for example, every 100 ms), the IoT gateway aggregates and groups the received data into a blockchain block and sends it to the cloud server. Note that each block can contain multiple transactions. For example, the readings of a certain sensor can be aggregated into a single transaction. Similarly, if the doctor is sending configuration commands to the IoT sensor, the configuration settings of each sensor can be grouped into a transaction. Each transaction will be signed by the owner that created the transaction.

\subsection{Consensus Protocol}

We consider a network of cloud servers that are used by various healthcare providers to manage the system. As mentioned, the cloud servers act as full blockchain nodes that store all the blockchain blocks. In addition, the cloud servers participate in the blockchain consensus protocol. Each cloud server has a unique blockchain ID. The cloud servers create the blockchain blocks successively based on their IDs. In other words, the server with the smallest ID creates the first block, followed by the server that has the second smallest ID, and so on. When the server that has the biggest ID creates a block, the turn goes back to the first server. Note that the block generation time at the IoT gateway should be adjusted to allow all the cloud servers to generate their blocks in order to avoid block accumulation at the cloud servers.

\begin{figure}
\centering
\includegraphics[width=1.0\textwidth]{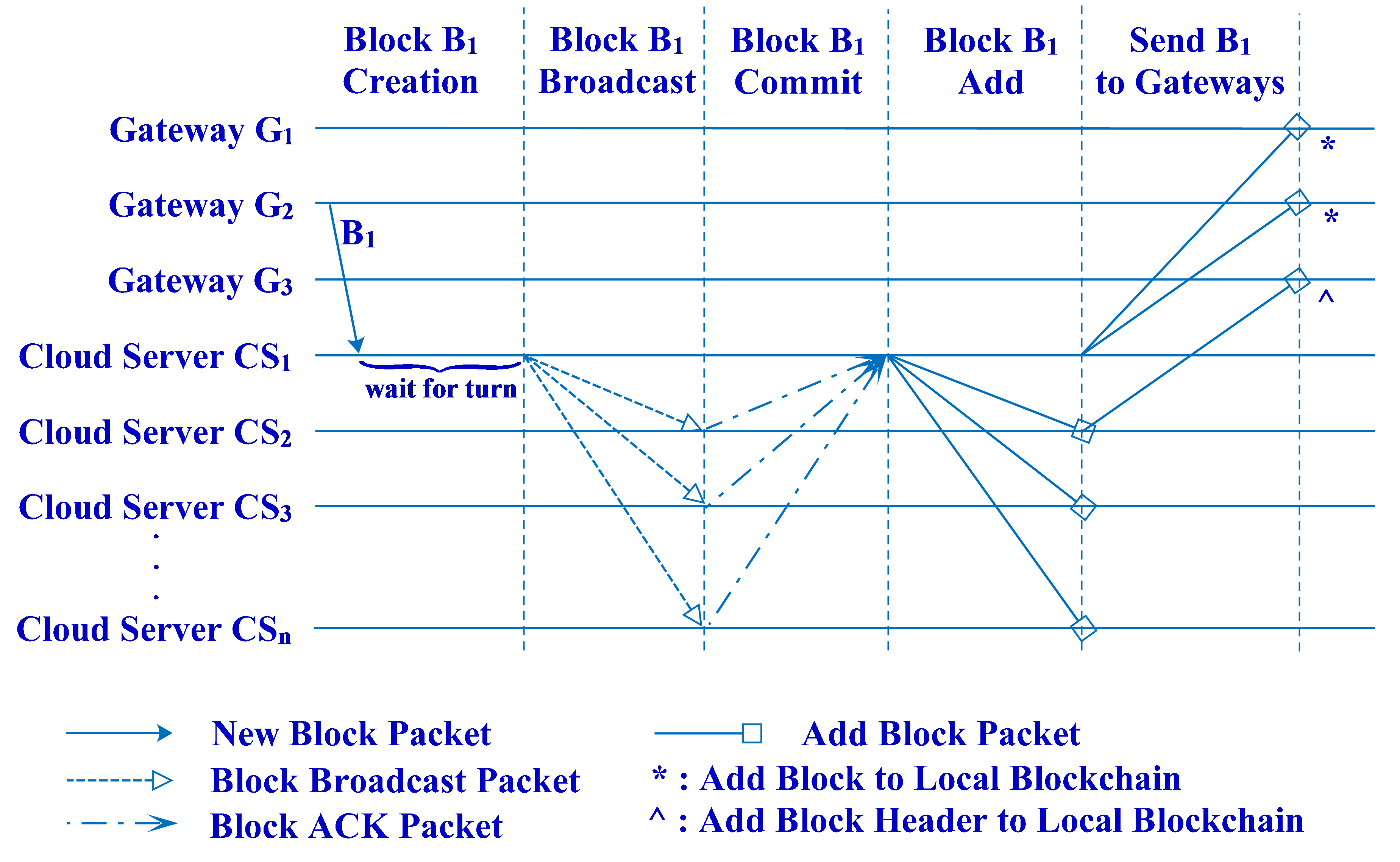}
\caption{A sample scenario of the proposed consensus algorithm.}
\label{fig:consensus}
\end{figure}

When its turn to create the new block arrives, a cloud server \textit{CS}\textsubscript{1} broadcasts the block that it received from the gateway to all the cloud servers. Each cloud server \textit{CS}\textsubscript{i} verifies that all transactions in the block are legitimate by validating the signature of each transaction. Next, \textit{CS}\textsubscript{i} replies with a CONFIRM message to \textit{CS}\textsubscript{1}. The confirm message contains \textit{CS}\textsubscript{i}'s signature of the new block. However, If \textit{CS}\textsubscript{i} discovers that one or more transactions in the block are not valid, it replies with an ERROR message. In its turn, \textit{CS}\textsubscript{1} waits until it receives at least (\textit{N}/2+1) CONFIRM messages before it adds the block to the blockchain and broadcasts its ID in a “Block Add” message to all cloud servers. Here, \textit{N} is the number of the cloud servers. This mechanism allows a cloud server to add the new block after the majority of cloud servers confirm its validity. The "Block Add" message contains the signatures that \textit{CS}\textsubscript{1} received in the CONFIRM messages. When a cloud server \textit{CS}\textsubscript{j} receives a "Block Add" message, it checks the attached signatures to ensure that more than \((\textit{N}\div2)\) cloud servers have validated and confirmed the new block, before adding it to its blockchain.

After receiving the “Block Add” message, each cloud server adds the new block to its blockchain and broadcasts it to its clusters. Note that each cloud server can serve multiple institutions and organizations, with each institution/organization having its own IoT cluster. Each gateway in a cluster examines the new block to determine if it contains transactions that were generated by one of the IoT networks in the cluster. If yes, the gateway stores the block in its local blockchain and sends it to the IoT devices that are connected to it. Each IoT device validates the block (by hashing it and comparing the result to the hash in the block header) and then stores the block header in the headers’ blockchain. Next, the IoT device caches the block body for a small period of time before deleting it. On the other hand, if the new block does not contain transactions that were generated by an IoT network in the cluster, the gateway validates the block, sends it to the IoT devices that are connected to it, extracts the block header, and adds it to the headers’ blockchain, and then deletes the block. Each IoT device that receives the block performs the same operations as the gateway. This allows the gateway and IoT devices to maintain the headers of all blocks in the blockchain and use these headers to validate any block from outside their local blockchain that they obtain from the cloud servers in the future.
The proposed consensus protocol is illustrated in Figure \ref{fig:consensus}. In the figure, gateways G\textsubscript{1} and G\textsubscript{2} are connected to cloud server CS\textsubscript{1}, while gateway G\textsubscript{3} is connected to cloud server CS\textsubscript{2}. At a certain time, G\textsubscript{2} creates a new block B\textsubscript{1} and sends it to CS\textsubscript{1}. When its turn to generate a new block arrives, CS\textsubscript{1} broadcasts B\textsubscript{1} to the cloud servers. Each cloud server confirms B\textsubscript{1} by sending a CONFIRM packet to CS\textsubscript{1}. Next, CS\textsubscript{1} sends a “Block Add” packet to the cloud servers, and each cloud server sends the new block to its gateways. G\textsubscript{1} and G\textsubscript{2} receive B\textsubscript{1} from CS\textsubscript{1} and add it to their copies of the local blockchain (since B\textsubscript{1} was generated by a gateway in CS\textsubscript{1}'s cluster), while G\textsubscript{3} receives B\textsubscript{1} from CS\textsubscript{2} and adds its header only to the local blockchain.

\subsection{Smart Contracts and Data Management}

When an IoT device or a user requires data from the blockchain, it sends a request to the IoT gateway. The latter searches for the data in its local chain. If it finds it, the gateway authenticates the sender and verifies that it has access to the requested data. If yes, the gateway replies directly to the sender with the block that contains the data and the token that enables the sender to access the data (more about this soon). If the gateway finds that the required data doesn't exist in the local blockchain, it forwards the request to the cloud server. The latter performs the same operation, i.e., it authenticates the requesting node and verifies that it has access to the requested data. If yes, the cloud server sends the block that contains the data and the access token to the gateway, which forwards them to the sender. When the latter receives the block, it validates it using the headers blockchain before it retrieves the required transactions from the block and decrypts it using the access token. 

\begin{figure}
\centering
\includegraphics[width=1.0\textwidth]{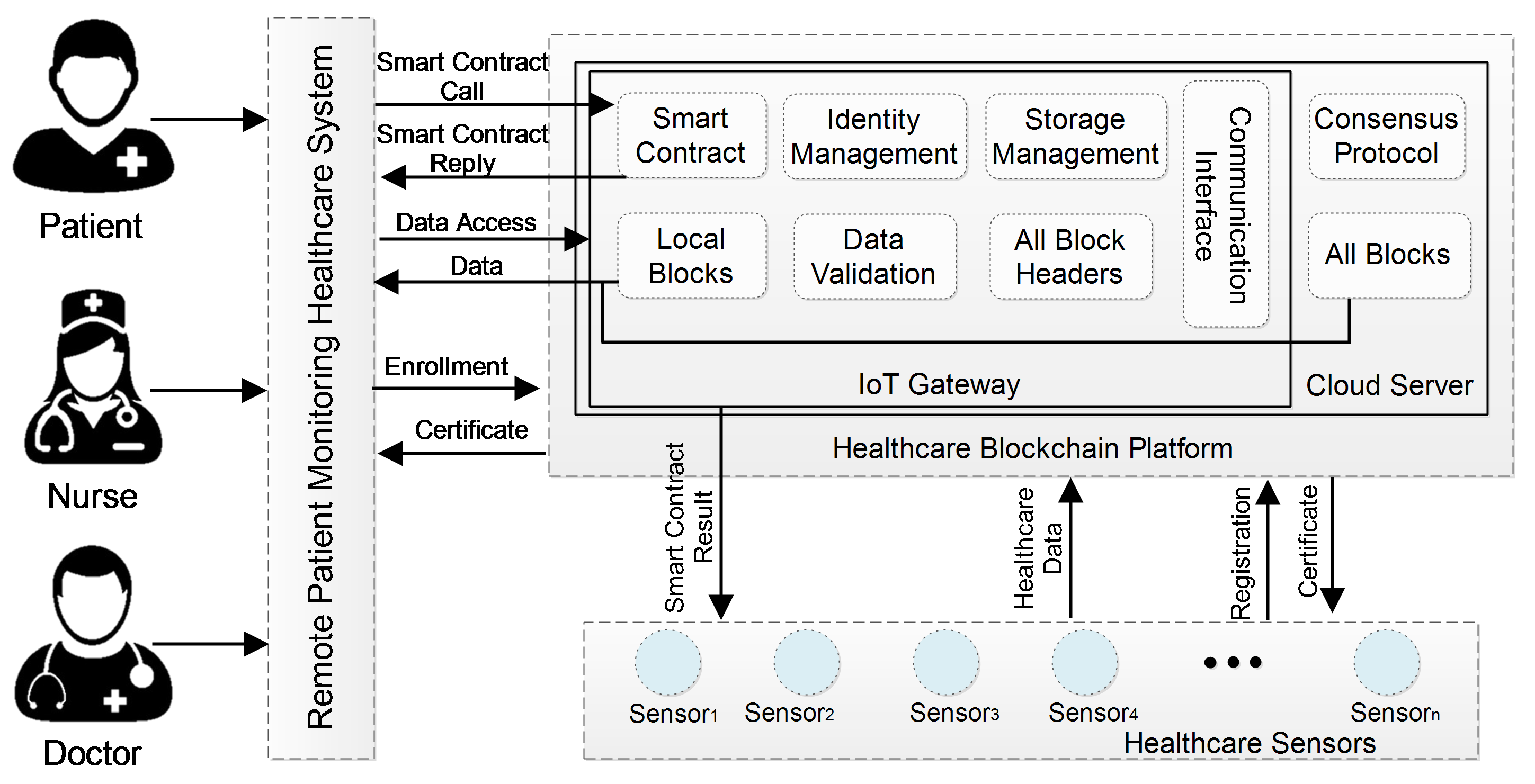}
\caption{Interaction Model of the proposed blockchain-based remote patient monitoring system.}
\label{fig:BCComms}
\end{figure}

Note that in our system, all transactions that can be accessed together are assigned an access token by the creator and saved into a smart contract. When the creator wants to grant access to the transaction to a certain node/user, the creator executes a smart contract function that adds the ID of the node/user to the access list of these transactions that is saved in the smart contract. When the node/user wants to access the transactions, it should authenticate itself and obtain the access token as described before. If the transactions belong to the local chain, the smart contract is executed by the gateway within the local chain. Else, the smart contract is executed by the cloud server within the full chain. The various subsystems and interactions in the proposed RPM platform are presented in Figure \ref{fig:BCComms}.

 \section{Specifications of Gateways at the Fog Layer}
\label{sec:fog}

The fog layer is made up of IoT gateways which function primarily as a hub between the cloud and IoT levels \cite{cite1}. With an in-depth study of the role of the gateway in a smart home/hospital, where the location and mobility of things and users are confined to hospital premises or buildings, it can be recognized that the stationary nature of the gateways empowers them with the property of being non-resource constrained in terms of power consumption, processing power, and communication. These advantages can be used by allowing gateways with ample intelligence, computing power, and structured networks.

 An inter-device communication is the key task of a gateway and supports numerous wireless protocols. We broaden the function of such gateways into fog enablers by (1) building a distributed gateway network and (2) implementing features such as the repository (i.e., local data processing and storage using blockchain) to temporarily preserve data for analysis by sensors and users. These are important to provide local pre-processing of sensor information and, therefore, to be an intelligent gateway for medical services. In a smart gateway, the main functions are:

\subsection{Local data processing and storage} Local data processing is a key aspect of fog computing and is performed locally so that intelligence is accessible at the doors. Based on the device architecture, fog/edge layers must continuously handle a large amount of information and respond to different conditions in a short time. In the remote patient management system, this becomes more important by allowing the system to respond to medical emergencies as quickly as possible. Gateways should store inbound information in local storage to ensure that the remote patient monitoring system can quickly recuperate patient medical data. In the proposed system, we make use of the local blockchain to achieve this objective in a secure manner. The patient data can be stored as blockchain transactions in an encrypted or compressed form depending on their context and security requirements. The gateway stores all data related to the local cluster in the local blockchain. In addition, when the gateway receives a blockchain block that contains data related to other clusters, it caches the block for a small period of time to allow users in the cluster who require data from the block to access it in a fast manner while the data is hot. % For other gateway activities, local data storage is also important. 
Moreover, since the network bandwidth is limited between the gateway and the cloud, the locally cached blocks can be used to maintain a continuous data flow in the event of a weak or unstable connection.  %incongruous data transmission and processing. When the network is not usable, data storage on gateways makes it reliable and secure.

\subsection{Data filtering} Data from various medical sensors must be obtained before further processing, e.g., data analysis, on the fog layer. The major sources of knowledge for the assessment of the health status of a patient in the remote patient monitoring system \cite{cite3, cite4} are bio-signals, for example, Electroencephalography (EEG), Electrocardiogram (ECG or EKG), and Electromyography (EMG). They typically have complex types that have small amplitudes and varying frequencies. It is important to remember that noise is often introduced to the signals during a patient's body sensing in a way that distorts the accuracy of the signals. These noises are caused by different sources, including electromagnetic interference from other electrical devices, shifts in current in the electricity grid, and inappropriate attachment of sensors to the body of users. At the fog level, due to the proximity of the sensors, the gateway addresses this issue. The fog layer is digitized via different contact protocols by sensors (e.g., 6LoWPAN, Zigbee, etc.). While sensors are able to perform lightweight filtering to eliminate certain noises during the data collection process, the fog layer offers more complex and robust data filtering.

\subsection{Data analysis} With local data analysis in the fog layer, the sensitivity of the device can be corrected. It helps the device to anticipate and diagnose situations of emergency. 
The developed deep learning module for detecting irregular cardiac conditions is implemented in the fog layer in our proposed RPM medical system. The deep learning module can categorize signals and detect abnormal conditions on the basis of the sensed ECG signal. 
As a result, the device responds more accurately, rapidly, and in real time to emergency situations. In addition, local input and locally sensed data analysis change the quality and reliability of the device in the event of the unavailability of the Internet link. Internet disconnection may occur regularly for the long-term monitoring of patients with chronic diseases. Fog computing, in this case, provides local maintenance of the system's features. Thus, the sensed data and processing results can be kept locally on the fog layer and later synchronized to the cloud via the blockchain. Data analysis in the fog often helps the device minimize severe parameter processing latencies.

\subsection{Improved latency} Agile responses and quick decision-making for acute diseases and emergencies, where transmission time and data processing are to be reduced, are important for a continuous remote control system. When raw medical data is transferred from medical sensor nodes to the cloud, cloud computing can trigger response latencies indefinitely if the network condition is not predictable. This becomes serious when streaming-based data processing, such as that EEG or ECG signals that are obtained from patients, is needed. Hence, deploying high-priority data analytics in distributed gateways in the fog later and making time-sensitive and critical decisions inside the local network make the remote patient monitoring system more predictable and robust. The processed data can then be transmitted for storage and further processing to the cloud.

\subsection{Sensor nodes energy efficiency} There are various drawbacks to the processing of data at sensor nodes, as medical sensors are resource-restricted devices. Complicated tasks can, in certain cases, be performed successfully at sensor nodes but at significant energy costs. The transfer of heavy-weight tasks from sensors to intelligent gateways in the fog layer can be an effective solution for solving the above-mentioned problem, in particular when sensors do not have sufficient resources. Much energy can be saved with the aid of fog computing by outsourcing tasks from medical sensors to intelligent gateways.

\section{Performance Evaluation}
\label{sec:perfEval}
In this section, we present the performance evaluation of the proposed RPM system. We have mainly evaluated the performance of the proposed blockchain module and demonstrated the efficiency of Fog computing in dealing with critical healthcare applications. 

\subsection{Blockchain Implementation and Performance Evaluation}

The proposed blockchain model was implemented via the Hyperledger platform. Hyperledger is an open-source development platform for blockchain applications. It has been widely used as an implementation platform by the research community and is considered a benchmark tool to evaluate the performance of the proposed approach against state-of-the-art approaches. For smart contracts, the Hyperledger tool provides easy-to-configure and user APIs, thus making validation easy for our research work. Furthermore, the RESTful API is utilized to provide the functionality of interoperability and expose the back-end blockchain services to the client application through which the patients or other medical personnel interact with the system. The smart contacts are designed and aggregated in the form of \verb|.bna| files known as business network archive. Hyperledger Composer \cite{Composer} is used to implement and design the proposed medical blockchain, which aims to enhance system operations in terms of throughput and latency. Hyperledger Composer is an open-source tool used to design blockchain applications. The \verb|.bna| in the designed platform consists of a model, query definition, transaction, and access control rules. The model file is the combination of participants, assets, and transactions. The participants are the user of the system who can interact with the system to commit transactions. Similarly, the assets are the medical services that are used by the system users (participants), which are stored in the blockchain. Likewise, transactions are operations that are used to communicate with assets. Moreover, transactions are also used to amend the values of assets and participants. Similarly, the access control rules are also defined to yield authentication and authorization to the users of the system. We also used the world state database to store the blockchain data. We specified the queries that are required to determine the interaction between the blockchain and the world state database. The queries are also used to fetch the user-based customized data from the database.

\autoref{Tab} encapsulates the business network archive file with transactions, assets, and participants. The users are patients, doctors, and nurses. Similarly, the assets comprise patients' medical records, sensors, vital sign readings, and other healthcare records. Lastly,  transactions include \verb|getVitalSignReadings|, \verb|AddHealthcareSensor|, and \verb|DetectStatus|.

\begin{table}[h!]
	\centering
	\caption{Smart Contract Modeling for Proposed RPM System}
	\label{Tab}
	
	\begin{tabular}{|p{2cm}|p{5cm}|p{6.5cm}|}
		\hline
		Type                         & \multicolumn{1}{c|}{Components} & \multicolumn{1}{c|}{Description}                                                                                                                                                                                             \\ \hline
		\multirow{3}{*}{Asset}       & Healthcare_Sensor                         & Healthcare sensors, such as ECG, or EMG etc. \\ \cline{2-3} 
		& Vital\_Sign_Sensing_Data           & The vital signs of patients acquired from healthcare sensor.                                                                                                             \\ \cline{2-3} 
		& HealthRecord                   & The patient medical information, such as current health condition, deployed sensors, etc.                                                                         \\ \hline
		\multirow{3}{*}{Participant} & Doctor                         & System user.                                                                                                                                                                                                          \\ \cline{2-3} 
		& Patient                        & System user.                                                                                                                                                                                                          \\ \cline{2-3} 
		& Nurse                          & System user.                                                                                                                                                                                                          \\ \hline
		\multirow{4}{*}{Transaction} & getVitalSignReadings     &Get vital sign reading from healthcare sensors.                                                                                             \\ \cline{2-3} 
		& Add_Healthcare_Sensor                     & Addition of new healthcare sensor in a medical blockchain platform.                                                                                                                                                                                         \\ \cline{2-3} 
		& Modify_Sensor                  & Modify sensor composition.                                                                                                                      \\ \cline{2-3} 
		& Detect_Status          & Detect the patient vital sign status.                                                                                     \\ \hline
	\end{tabular}
\end{table}

\begin{figure}[!t]
    \centering
  \includegraphics[scale=0.21]{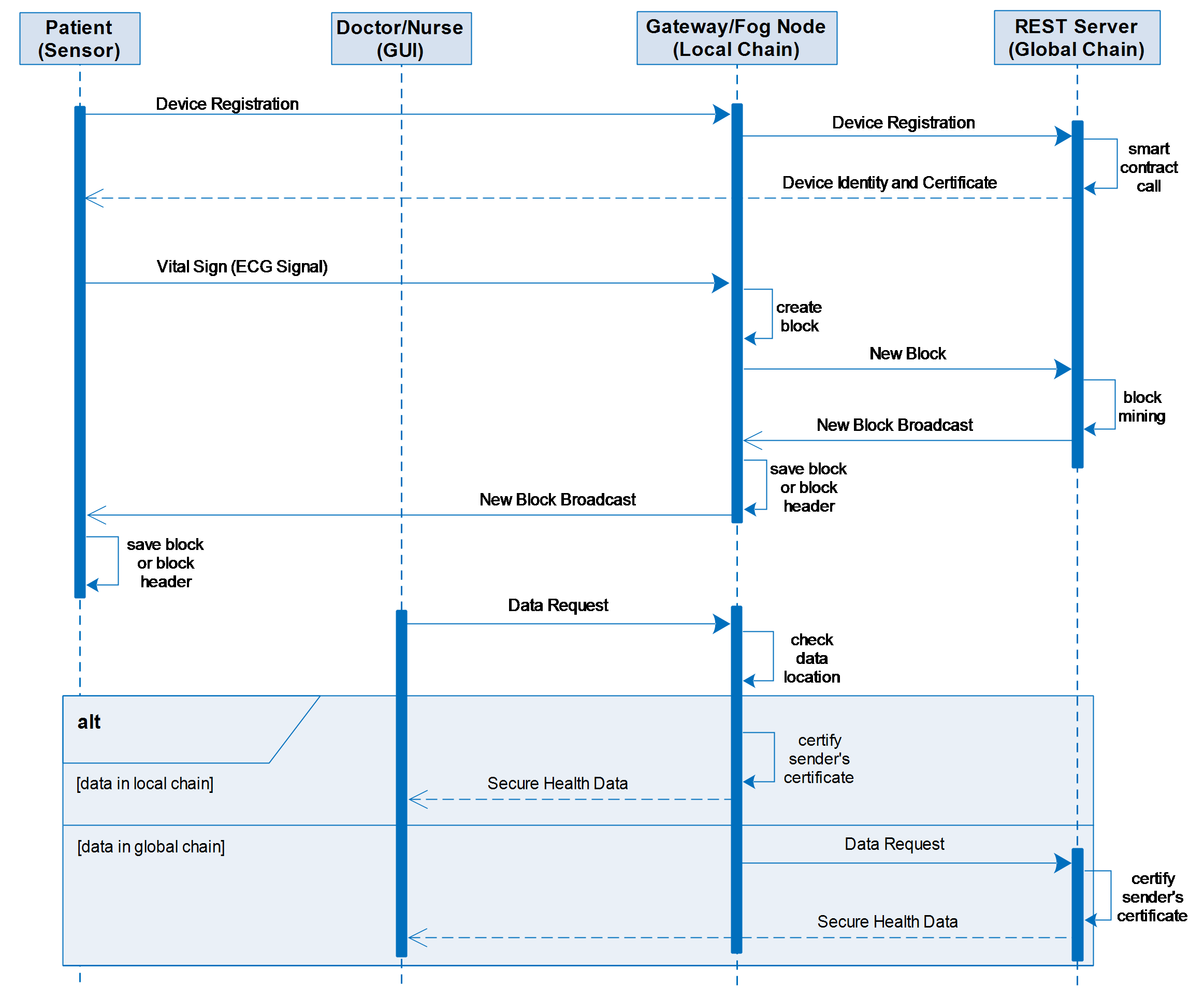}
    \caption{Sequence of interactions conducted during simulation}
    \label{fig:interaction}
\end{figure}

The business network archive is then used to construct a Representational state transfer (REST) Application Program Interface (API) in order to provide communication between the client application and the back-end database. The RESTful API provides cross-accessibility, where the user of the system can access it from any platform with authentic credentials. \autoref{rest1}, presents the RESTful API for the proposed medical blockchain, which is based on HTTP protocol. The generated RESTful API is used to expose the medical platform services to the client application. The services are related to patients, nurses, doctors, EMR, and other medical information. 
% \color{red}
Figure \ref{fig:interaction} demonstrates how the major components of the proposed RPM system have interacted during the simulation study.  
\color{black}

\begin{table}[h!]
		\caption{RESTful API for proposed Medical Blockchain}
	\label{rest1}
	\begin{adjustbox}{max width=\textwidth}
	\begin{tabular}{llll}
		\hline
		\textbf{Action}                                   & \textbf{Verb} & \textbf{Media Type} & \textbf{URI}                \\ \hline
		Patient Dashboard                                & ALL           & Application/json    & /api/Patient                \\
		Doctor Dashboard                                 & ALL           & Application/json    & /api/Doctor                 \\
		Nurse Dashboard                                  & ALL           & Application/json    & /api/Nurse                  \\
		Healthcare Sensor Dashboard                                   & ALL           & Application/json    & /api/Sensor                   \\
		Vital_Sign    & Application/json    & /api/VitalSignReading       \\
		EMR Dashboard & ALL     & Application/json    & /api/PatientRecord       \\
		Share patient record with healthcare personnel & POST     & Application/json    & /api/ShareRecord       \\
		Blockchain Network Text                           & GET           & Application/json    & /api/system/ping            \\
		Issue identity to system user        & POST          & Application/json    & /api/SystemIdentities/issue \\
		Get Identities                           & GET           & Application/json    & /api/System/identities      \\
		Retrieve historian records                         & GET           & Application/json    & /api/System/historian       \\ \hline
	\end{tabular}
	\end{adjustbox}
\end{table}

      % how to use Blockchain for the security of sensor readouts
    
%     how to use Blockchain for patient access control (doctors or any caregivers)
    
%     security analysis: how Blockchain can fulfill security requirements of RPM system.

Within our blockchain implementation, each piece of medical record has one user (owner) who can share the data they own with other users (doctors) at varying levels of access. Data sharing between users is modeled by a system where users can share data with other users in different groups, as well as receive data requests from other users at any access level. If a user responds to a request by granting data access, an access token is provided to the receiver in a way that allows that receiver to access the data at the specified access level only. Our system ensures that sensitive information is never exposed on the blockchain, including both private and document keys, which is necessary in order to maintain the privacy and security of user-controlled data.

\begin{figure}[H]
    \centering
    \includegraphics[scale=0.5]{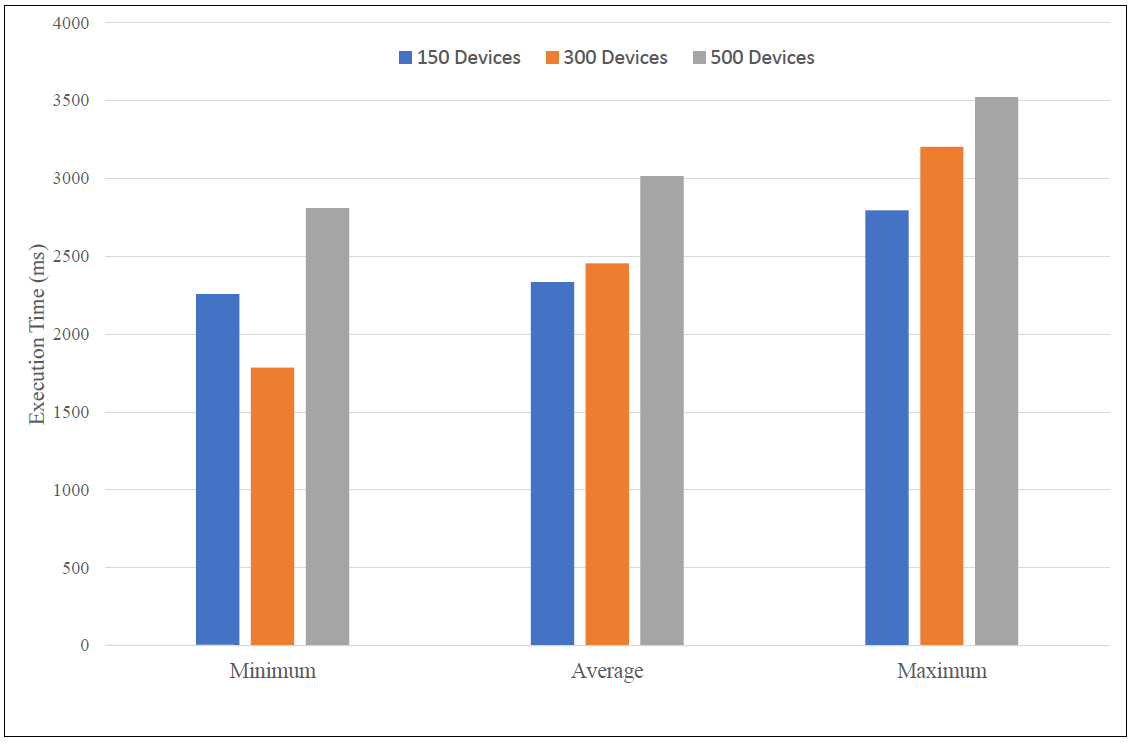}
    \caption{Healthcare device registration execution time}
    \label{fig:regExecTime}
\end{figure}

% \subsection{Blockchain module performance evaluation}
We evaluate the performance of the proposed blockchain model using Hyperledger Caliper \cite{caliper}. For experimental analysis, we carried out several experiments in terms of the execution time when adding a new healthcare device and executing a healthcare data query. We also measure the average time of the proposed consensus algorithm. The execution time is the round-trip time which includes the total time of sending the request by the client and getting the response from the network. In order to evaluate the execution time, we utilized the Postman tool, which is used to explore and test the RESTful APIs by simulating a customized user load within the network. In this study, we created three groups of devices: 150, 300, and 500, in order to investigate the execution time of registering a device in the proposed blockchain model. Furthermore, the execution time is analyzed using different statistical measures, such as the minimum, maximum, and average times. As shown in \autoref{fig:regExecTime}, in the case of 150 users, the average, minimum, and maximum execution time to register the healthcare device is recorded as 2335 ms, 2257 ms, and 2795 ms, respectively. Likewise, the minimum, maximum, and average execution times for 300 healthcare device-group is are 1785 ms, 3204 ms, and 2454 ms, respectively. Finally, for 500 devices the minimum execution time is recorded as 2810 ms, whereas the maximum and average execution time is 3524 ms and 3015 ms respectively (\autoref{fig:regExecTime}).

\begin{figure}[H]
    \centering
  \includegraphics[scale=0.45]{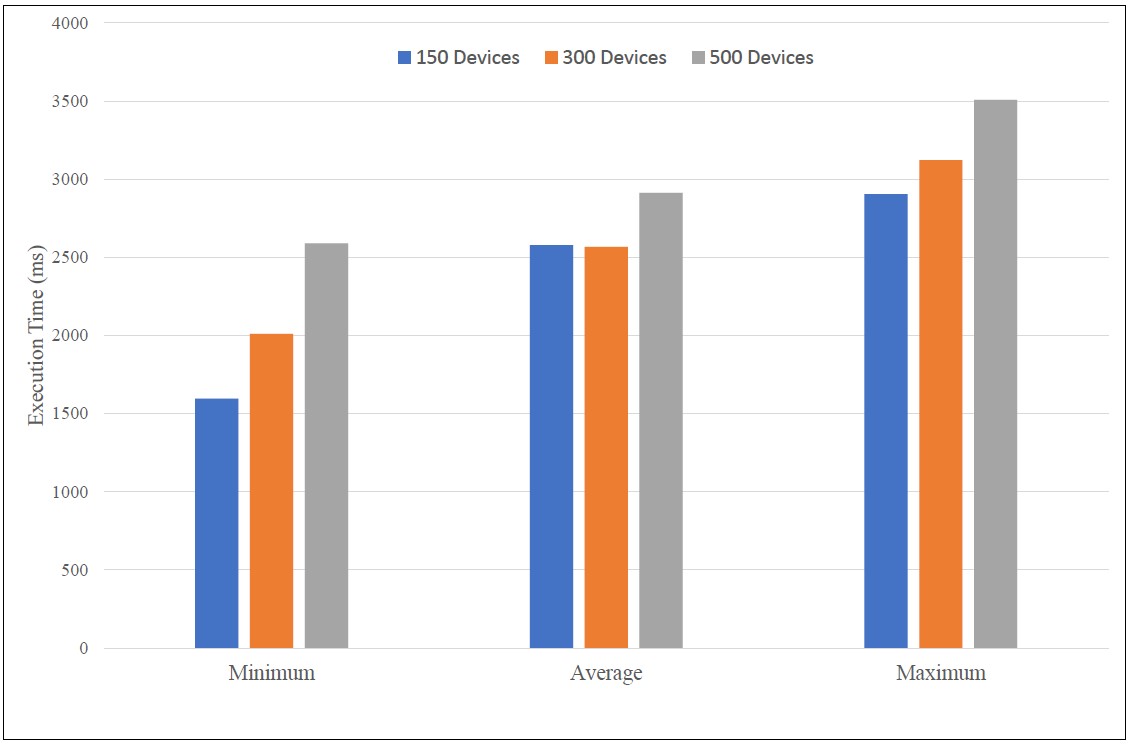}
    \caption{Vital signs reading execution time}
    \label{fig:TimeExec}
\end{figure}

The execution time of the proposed system is also evaluated in the case of retrieving healthcare data from the blockchain network. Every healthcare device in the proposed platform has the HTTP client functionality which is used to send requests for vital sign sensing data through the IoT gateway. The request is initially processed by the IoT gateway. If the requested data is found in the local chain, the IoT gateway validates the device certificate via the local smart contract and then replies to the device with the encrypted data. Else, the IoT gateway forwards the request to the REST server, which performs a similar process. %Since the vital sign data is stored in the Blockchain, the REST server is used to retrieve the data and expose it to the client application based on the request. 
The execution time of reading the vital sign data is illustrated in Figure \ref{fig:TimeExec}. The same set of device groups has been considered for the experimental evaluation, i.e., 150, 300, and 500 devices. It is observed from the graph that the increase in the device scale in the proposed healthcare system will also create an impact on the execution time. However, the overall execution time of the network remains stable until there is high congestion in the network. 
%In case of Ethereum and Bitcoin, it requires 15 seconds to 2 minutes and 10 minutes to an hour, respectively, to mine a new block. The execution time to mine a block in the proposed platform is based on the network environment. 
The average execution time of vital sign sensing data in the case of 150, 300, and 500 devices are 2552 ms, 2525 ms, and 2775 ms, respectively, which are comparable to the execution times of registering a device that is shown in \autoref{fig:regExecTime}. %Hence  the proposed permissioned Blockchain network outperforms other Blockchain platforms in terms of execution time.

%Khaleel - to do: add a figure for the consensus time of the proposed systems as compared to the consensus times of Bitcoin and Etherium 

\begin{figure}[H]
    \centering
  \includegraphics[scale=0.42]{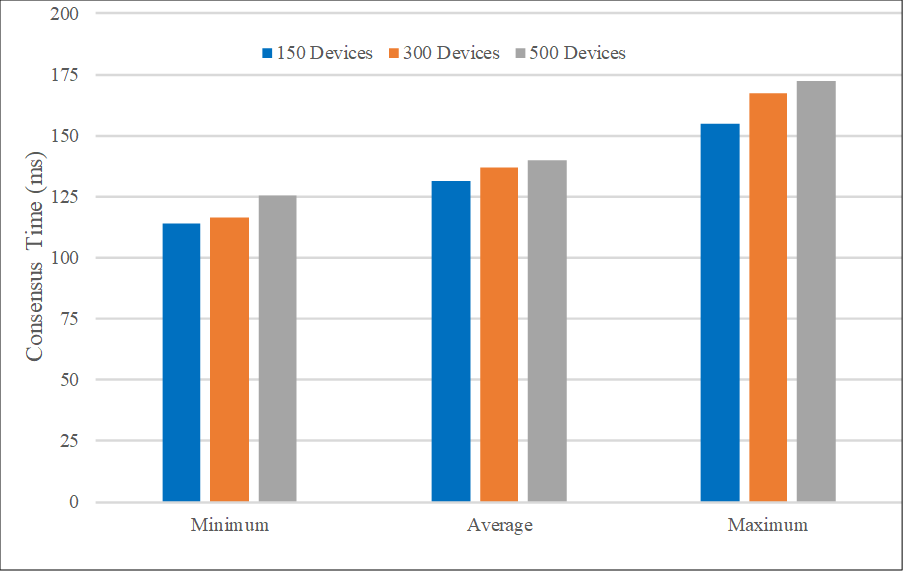}
    \caption{Block consensus time}
    \label{fig:TimeCons}
\end{figure}

In order to evaluate the effectiveness of the proposed consensus method, we tested several scenarios in which we deployed five REST servers and five IoT gateways. The IoT devices were distributed evenly among the gateways, and each gateway was connected to a REST server. The servers saved all the blocks that were confirmed by the consensus protocol, while the IoT gateways saved the blocks of the devices that connected to them only. In these scenarios, we measure the consensus time of each created block, then we calculate the minimum, maximum, and average values for all the created blocks. The results are shown in Figure \ref{fig:TimeCons}. We notice that the consensus time generally increases as the number of devices increases, which is logical since, with more devices, the total number of transactions increase, which adds more time to validate the new blocks. However, the increase in the consensus time is only 12.5 ms (on average) as the number of devices increases from 150 to 500, which proves the efficiency of the proposed consensus approach. In addition, the average consensus time of the system is 140 ms. In case of Ethereum and Bitcoin, it requires 10 to 19 seconds and 10 minutes to an hour respectively to mine a new block. Hence, the proposed consensus algorithm outperforms those of other blockchain platforms in terms of consensus time.

\subsection{Efficiency of the Fog computing infrastructure}

\begin{figure}[!t]
    \centering
  \includegraphics[scale=0.50]{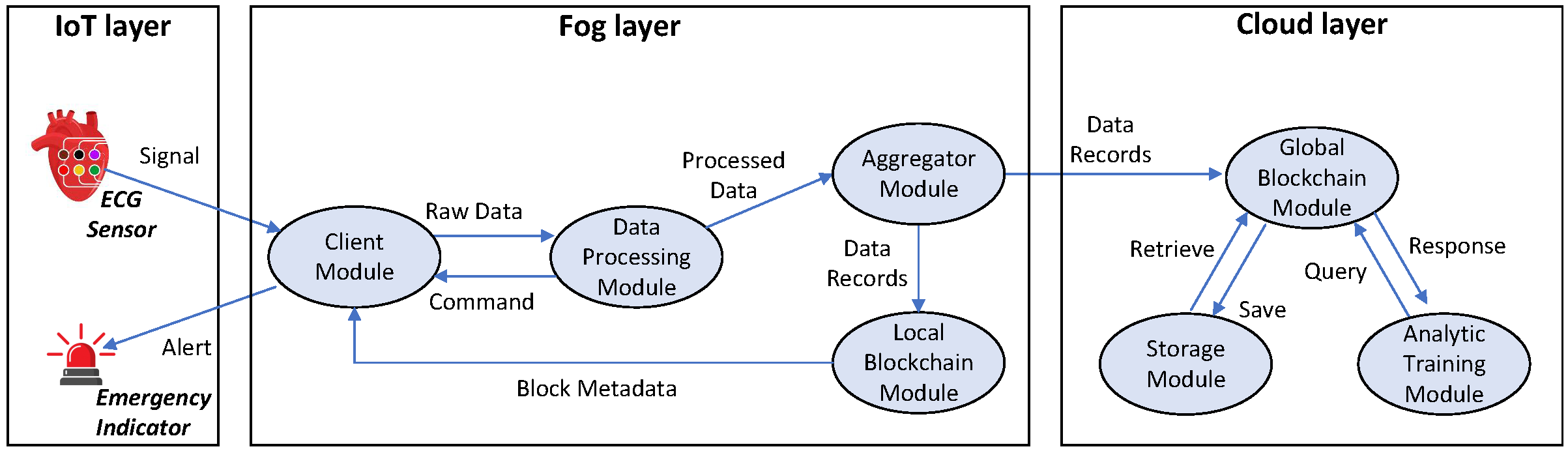}
    \caption{Data flow model for the proposed RPM system}
    \label{fig:appflow}
\end{figure}

Figure \ref{fig:appflow} depicts the distributed data flow model for our proposed IoT-driven critical healthcare applications. According to this model, data signals generated by the IoT devices are pushed into the client module, an initial application interface for interacting with the IoT devices and actuators and receiving the user's information, such as name, location, address, sex, and age of the patient. After pre-processing and filtering the data that is coming from the IoT devices, the client module forwards the data to the Data Processing module for further processing. Here, AI-enabled modules can execute data analytics processes for testing purposes. Based on the outcome of the data processing, a command is issued by the Data Processing module for the client module so that it can trigger physical emergency actions through the actuators. Next, the Data Processing module dispatches the processed data to the aggregator module, which simultaneously interacts with the blockchain module at the IoT gateway and cloud server to add the data to the blockchain and ensure data integrity and location-independent data access. The blockchain module interacts with the storage module in case the data is to be stored off-chain. Finally, The Data Processing module at the cloud server interacts with the blockchain module to consistently produce the results that are requested by the application users. %Once the model is trained with new data, the model is replicated to the data analytics module for testing with new data. Furthermore,  
Since the client module directly interacts with the IoT devices, it is preferable to be deployed at the IoT gateways (e.g., ECG machines). For the deployment of other modules, there exist different approaches in the literature. For instance, cloud computation has been exploited in \cite{reference1} \cite{reference2} to execute the data analytics, aggregator, blockchain, storage, and training module. On the other hand, the proposed RPM system adopts Fog computing for executing these modules and utilizes the cloud to host the blockchain, storage, and processing modules. 

\begin{figure}[!b]
    \centering
  \includegraphics[scale=0.75]{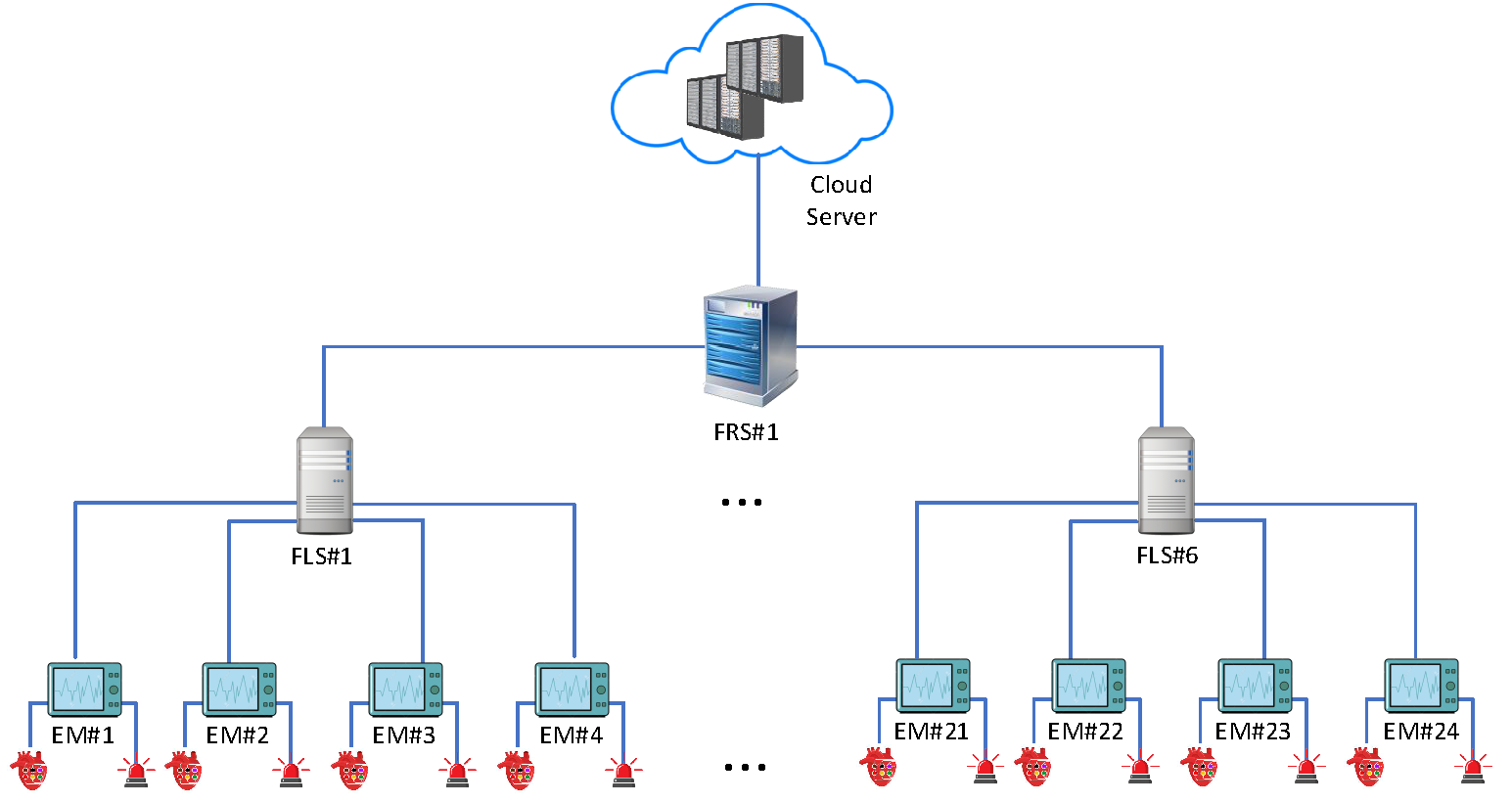}
    \caption{Architecture of the simulated Fog-Cloud computing environment}
    \label{fig:simenv}
\end{figure}

In this phase of performance evaluation, we demonstrate how the augmentation of Fog computing in remote patient monitoring improves the service latency and the energy usage in comparison to harnessing cloud-based resources. The experiments are conducted in an \textit{iFogSim} \cite{iFogSim} simulated Fog-Cloud computing environment. The computing resources within the simulation environment are organized in a hierarchical order, as shown in Figure \ref{fig:simenv}. At the lower level of the simulation environment, twenty-four ECG machines (EMs) equipped with ECG sensors and emergency alert systems are placed. Based on the simulation design, an EM can connect with any of the four Fog local servers (FLSs) at the upper level. All FLSs are also set connected with a Fog regional server (FRS) that helps the lower-level computing devices to maintain seamless communication with the Cloud datacenter. Table \ref{tab:simulation} presents the details of the simulation parameters used in the experiments. The numerical values have been extracted from real-world references as specified in \cite{EdgeAffinity} \cite{HAL}. Additionally, Table \ref{tab:TupleConf} illustrates the configuration of different application modules for the simulations, which have been approximated based on the profiled run-time, resource utilization, and data communication delay of the proposed solutions in heterogeneous computing devices and networking context.  

\begin{table}[!t]
\centering 
\caption{Parameters of simulated environment}\label{tab:simulation} 
\small
\begin{tabular}{p{1 cm}p{2 cm}p{2 cm}p{2 cm}p{2 cm}p{2 cm}p{2 cm}}
\hline
\multicolumn{7} {p{4cm}}{Device configuration}\\
\cline{1-7} 
Name & Processing \newline speed &  Downlink \newline bandwidth &  Uplink \newline bandwidth & Memory \newline capacity & Busy \newline power & Idle \newline power\\
     & (in MIPS) & (in MB) & (in MB) & (in GB) & (in MWh) & (in MWh)\\\hline
EM & 1000 & 10 & 5 & 8 & 1.1 & 0.2 \\
FLS & 7000 & 8 & 3 & 12 & 1.3 & 0.4 \\
FRS & 15000 & 6 & 2 & 16 & 1.6 & 0.8 \\
Cloud & 40000 & 3 & 4 & 32 & 3.2 & 1.4 \\ \hline 
\multicolumn{4} {p{6cm}}{Sensing frequency of ECG sensors} & \multicolumn{3} {p{4cm}}{5 signals per second} \\
\cline{1-7} 
\multicolumn{4} {p{6cm}}{Simulation time} & \multicolumn{3} {p{4cm}}{500 seconds}\\
\cline{1-7} 
\end{tabular}  
\end{table}     

\begin{table} %[ht]
    \caption{Module configuration}
    \centering 
    \label{tab:TupleConf}
    \begin{tabular}{p{6cm}p{2.8cm}p{2.8cm}p{2.8cm}} 
      Name & Program size  & Packet size & RAM usage\\
      & (in MB) & (in KB) & (in GB) \\
      \hline
      Client module & 2000 & 500 & 1\\
      Data analytic module & 4000 & 1500 & 6\\
      Aggregator module & 1500 & 1800 & 2\\
      Blockchain module (periodic) & 1000 & 2000 & 4\\
      Storage module & 1000 & 2000 & 2\\
      Analytic training module & 8000 & 2000 & 12\\
      \hline
    \end{tabular}
\end{table}

\begin{figure}[!t]
    \centering
  \includegraphics[scale=0.40]{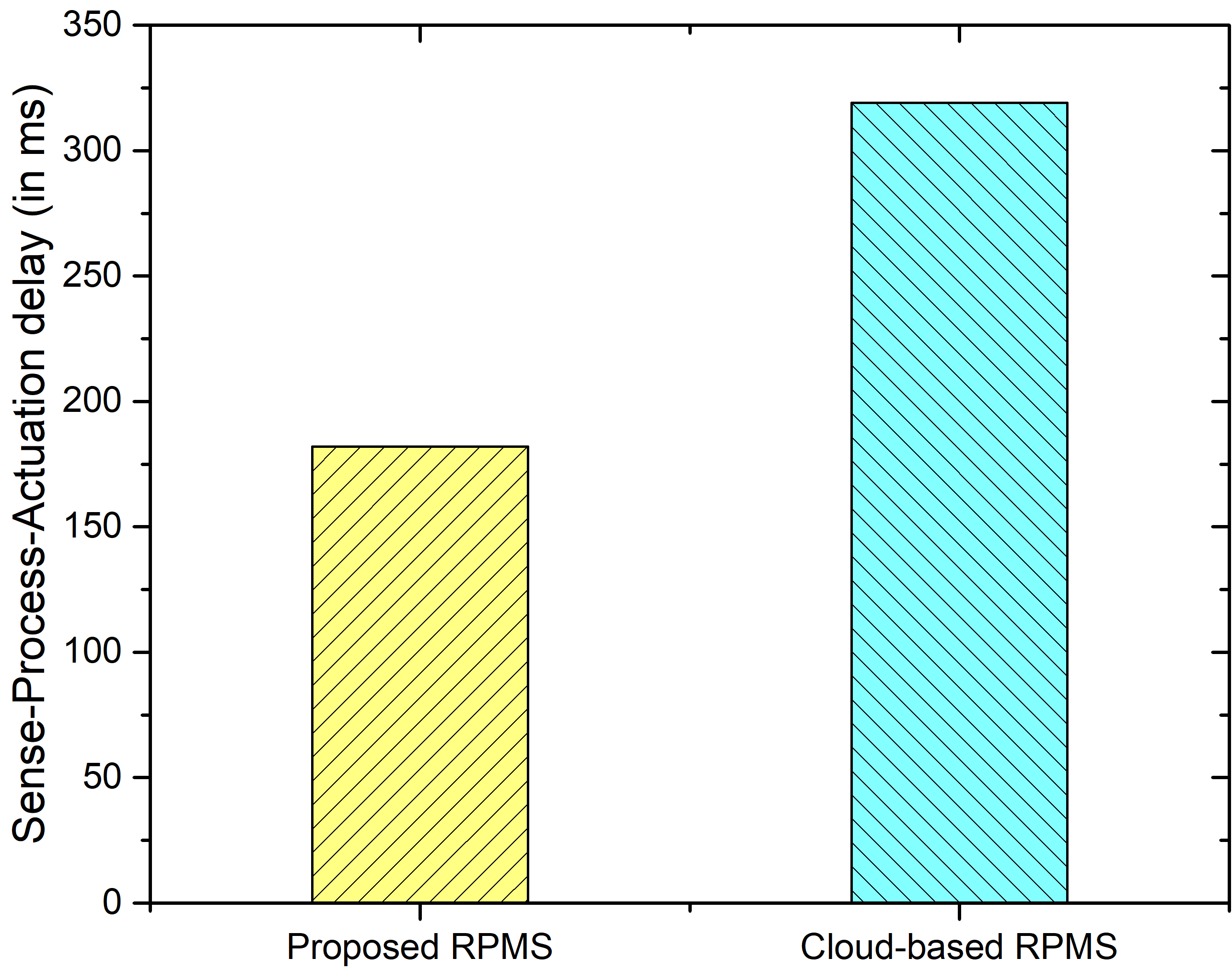}
    \caption{Performance in reducing sense-process-actuation delay}
    \label{fig:CycleDelay}
\end{figure}

\begin{figure}[!t]
    \centering
  \includegraphics[scale=0.40]{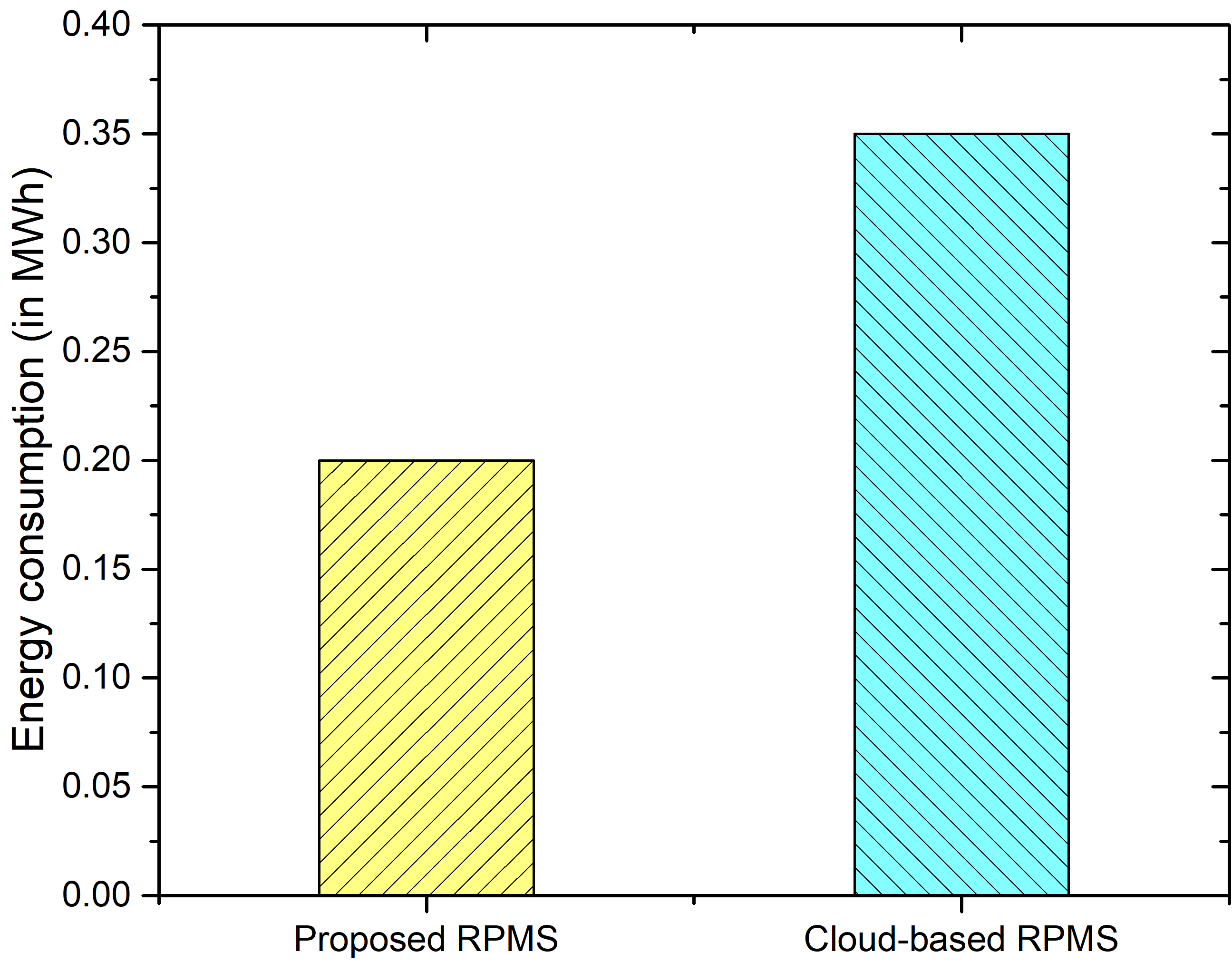}
    \caption{Performance in reducing energy consumption}
    \label{fig:EnergyCon}
\end{figure}

\begin{figure}[!t]
    \centering
  \includegraphics[scale=0.330]{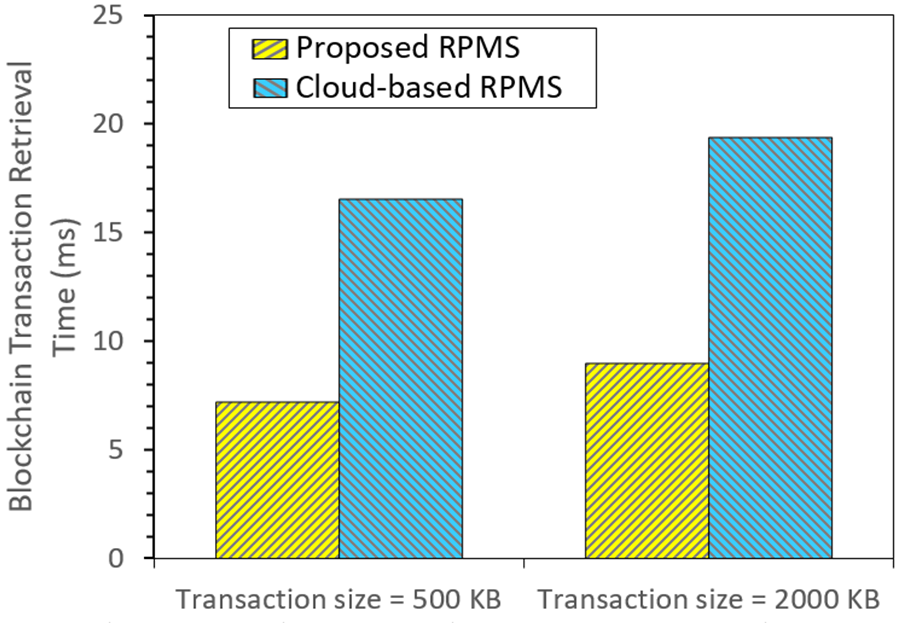}
    \caption{Performance in blockchain transaction retrieval}
    \label{fig:Blockchain}
\end{figure}

The results of the simulation experiments conducted in the aforementioned computing setup demonstrate that our proposed Fog computing-based RPMS outperforms the Cloud computing-based RPMS both in terms of reducing sense-process-actuation delay (calculated using iFogSim AppLoop model on ECG sensors $\rightarrow$ client module $\rightarrow$ data analytic module $\rightarrow$ client module $\rightarrow$ emergency alert system data flow) and energy usage. Figure \ref{fig:CycleDelay} indicates that the augmentation of Fog computing can improve the responsiveness of RPMS by 40\% in initiating alert messages during emergency situations compared to its cloud counterpart. Such performance improvement happens mainly for executing the data analytics module closer to the sources, that consequently decreases the data transfer delay to remote cloud servers. Moreover, the computing devices in the Fog paradigm consume a reduced amount of energy than a cloud server because of their capacity constraints. Statistically, this feature also has an influence in lowering the idle energy consumption of Fog computing devices. Therefore, when the time-based energy consumption model (as programmed in the iFogSim simulator) is applied, the Fog computing-based RPMS promises to deliver its services by consuming around 36\% less energy than its Cloud-based implementation (as shown in Figure \ref{fig:EnergyCon}). 

On the other hand, due to executing the blockchain module at the fog devices, the delay required to retrieve a random blockchain transaction decreases as compared to the cloud-based RPMS, as shown in Figure \ref{fig:Blockchain}. The figure illustrates that when the transaction size is equal to 500 KB, the proposed system requires an average of 7.16 ms to retrieve the transaction from the blockchain, while cloud-based RPMS needs 16.54 ms. On the other hand, for a 2000 KB transaction, the proposed RPMS produces a delay equal to 8.9 ms, while the cloud-based RPMS needs 19.34 ms. Hence, the proposed RPMS reduces the transaction retrieval delay by an average of 55.1\%. This is mainly due to the cases in which the transaction is fetched from the local chain, which require much less end-to-end delay than retrieving the transaction from the global chain, due to the deployment of fog nodes at locations that are much nearer to the sensor nodes than the cloud servers. % (the block mining process) in a comparatively powerful physical resource than Fog infrastructure, for small size data in the transaction (data record packet size = 500KB), the Cloud-based RPMS performs slightly better in managing the Blockchain operation compared to the proposed RPMS as shown in Figure \ref{fig:Blockchain}. However, as the transaction size increases (data record packet size = 2000KB), the proposed RPMS outperforms the Cloud-based RPMS. This result bears testimony that, despite having lower-scale computing resources than the Cloud, Fog computing-based solutions can offer better performance by reducing data transmission time significantly, especially while dealing with a large amount of data.  

\section{Security Analysis}
\label{sec:security}
Having a robust architecture encryption scheme as part of a blockchain-based data-sharing system is particularly critical from a security perspective because most blockchain implementations replicate the entire transaction ledger onto each node, therefore, multiplying the potential attack surface by the number of nodes in the network. In the following, we discuss the security analysis which we performed on the proposed patient monitoring system.

\begin{itemize}

\item  \textbf{Key attack:} Elliptic curve encryption method is employed from a key pair, and an attacker can't calculate the private key to address the elliptic curve logarithm problem; hence the security of the proposed model is ensured. Moreover, for each session, a temporary private key is generated for interaction among the nodes. In such a way, if a private key gets compromised in terms of leakage, then this will not have an impact on the session, as the attacker would not be able to calculate a session key for a session that is currently going on among the nodes; and (b) the leaked private key is of no use until the session is completed. 

\item  \textbf{Replay attack:} The proposed model uses an individual temporary private key that is different for each session agreement among the interacting nodes. It is improbable that a replay attack becomes successful since private keys hold a bounded lifetime. 
\item  \textbf{Impersonation attack:} This attack is executed only if the attacker has successfully obtained the private key. The proposed model employs an individual private key and elliptic curve encryption. Therefore, this attack cannot be executed.

\item  \textbf{Sybil attack:} there are different methods to remove the impact of Sybil attack on the proposed model, such as increasing the price to form a new identity. This method restricts attackers from obtaining fake identities, using a two-factor authentication mechanism and accumulating the MAC and IP addresses of the participants, which permits the detection of those participants who have varying identities.

\item  \textbf{False data injection attack:} Prior to validating the records, the consensus algorithm is executed by the blockchain nodes. On arrival of the positive consensus, a node can confirm the legitimacy of the received record. 

\item  \textbf{Tampering attack:} For encryption and signing the transaction, a public key crypto-system is employed. This indicates that the tampering node cannot amend the transaction as it does not hold the private key of the signing node. Furthermore, the proposed model can handle the key attacks; therefore, the adversaries cannot exploit the private keys. 

\item  \textbf{Modification attack:} As explained above, this attack is impossible because the adversaries cannot exploit the private keys.

\item \textbf{Hiding blocks attack:} A record in the proposed vital sign monitoring platform holds a unique sequence number. It is a must for a blockchain node to provide its saved records if requested. If a node in the network does not offer its records, it is detached from the network and disallowed to interact with other nodes. 
\item \textbf{Man-in-the-middle attack:} A mutual authentication is performed between the nodes in the proposed model, which employs private keys for each session agreement, therefore, man-in-the-middle attacks are prevented.
\item \textbf{Compromisation attack:} If an attacker compromises a cloud server and attempts to sabotage the consensus operation by sending a "Block Add" message that contains an invalid block, the legitimate cloud servers will detect the attack from the invalid signatures in the "Block Add" message, since the attacker will not be able to generate the valid signatures of the other cloud servers. If the attacker drops the block that it receives from the IoT gateway, the latter reports the attack to the IoT ecosystem administrator when it detects that its block was not added to the blockchain in due time. Finally, if the attacker sends a wrong reply message when it receives a new block from another cloud server, the attack will not have an effect as long as the number of legitimate cloud servers is greater than \textit{N}/2. 
\end{itemize}

\section{Conclusion and Future Work}
\label{sec:concl}
In this work, we have presented a three-layer remote patient monitoring system that leverages blockchain technology for better security and Fog technology for providing low-latency services to IoT devices and healthcare users. The most important functions that encompass the system components are described and evaluated. In addition, a new consensus protocol that is tailored to the RPM environment is discussed and analyzed. 
Moreover, the blockchain module was implemented and tested using Hyperledger Fabric Framework, and it achieved low execution and consensus delays. %high throughput. 
Moreover,  the augmentation of Fog computing can improve the responsiveness of the remote patient monitoring system by 40\%.

% \color{red}
Several future works are being studied to enhance the proposed system. For example, we are planning to perform the simulations using real healthcare datasets (such as that in \cite{heart:kaggle}). In addition, we intend to add a prediction module at the cloud layer that can predict a heart disease problem before its occurrence. The module would analyze the patient's data from the global blockchain over an extended period to enhance prediction accuracy. Another enhancement would be the integration of the proposed blockchain system with a body area network (BAN) framework that is used to collect patient medical data in an efficient manner. Such integration should be carefully designed in order to secure the BAN operations without adding significant overhead in terms of computation and energy consumption on the BAN nodes. A similar system was proposed in \cite{shahbazi2020towards}. Hence, we aim to study the literature in order to adjust the proposed blockchain system to make it suitable for a BAN environment.

Another important future work is to enhance the proposed fog layer by augmenting it with modern technological tools that will improve its performance. For example, federated learning can be used by fog nodes to filter and analyze the readings of IoT devices in order to provide more accurate results to healthcare providers. Another important aspect is to design the scheduling of IoT data on the fog layer using the blockchain. For this aspect, we intend to adopt a previous strategy that we proposed in \cite{mershad2012score} to guarantee that a fog node treats data from IoT nodes fairly and provides equal opportunities for IoT nodes to save their data in the blockchain.  

Finally, we will study the scalability of the proposed system and its ability to support a large number of IoT ecosystems. For this purpose, we will design a hierarchical clustering framework that distributes cloud servers, fog nodes, and IoT devices into clusters based on their geographic locations and the deployed healthcare application. Using clustering will allow us to reduce the delay overhead when the application contains a huge number of blockchain nodes. In such a system, it is possible to execute a blockchain query in parallel by distributing it over the cluster heads, which would result in a reduced end-to-end delay between the patient and the healthcare provider.   
\color{black}

%Moreover, the system can be developed for other use cases.

% \section*{Acknowledgements}
% Dr. Omar Cheikhrouhou thanks Taif university for its support under the project Taif University Researchers supporting project number (TURSP-2020/55), Taif university, Taif, Saudi arabia.

\bibliographystyle{elsarticle-num}
% \bibliographystyle{elsarticle-harv}
% \biblio.graphystyle{elsarticle-num-names}
\bibliography{references}
% \end{enumerate}

\end{document}